\documentclass[final,5p,times,twocolumn]{elsarticle}

\usepackage{textcomp,graphicx,times,wrapfig,xspace,url,amssymb,amsmath,amsfonts,amsthm,amsbsy,wasysym,color,comment}
\usepackage[super]{nth}

\renewcommand{\[}{\begin{equation}}
\renewcommand{\]}{\end{equation}}
\newcommand{\eg}{{e.g.}\xspace}
\newcommand{\ie}{{i.e.}\xspace}
\newcommand{\cf}{{c.f.}\xspace}

\journal{Astropart. Phys.}

\begin{document}

\begin{frontmatter}

\title{Design constraints on Cherenkov telescopes with Davies-Cotton reflectors}

\author{T. Bretz and M. Ribordy} 
\address{High Energy Physics Laboratory, EPFL, CH-1015 Lausanne, Switzerland}

\cortext[corb]{Corresponding author: thomas.bretz@epfl.ch}

\begin{abstract}
This paper discusses the construction of high-performance
ground-based gamma-ray Cherenkov telescopes with a Davies-Cotton reflector.
For the design of such telescopes, usually physics constrains the field-of-view,
while the photo-sensor size is defined by limited options. Including the
effect of light-concentrators in front of the photo sensor, it is demonstrated
that these constraints are enough to mutually constrain all other
design parameters.
The dependability of the various design parameters naturally arises
once a relationship between the value of the point-spread functions at
the edge of the field-of-view and the pixel field-of-view is
introduced. To be able to include this constraint into a system
of equations, an analytical description
for the point-spread function of a tessellated Davies-Cotton reflector is
derived from Taylor developments and ray-tracing simulations. Including higher order terms renders the result precise on the percent level.\\
Design curves are provided within the typical
phase space of Cherenkov telescopes. The impact of all design parameters
on the overall design is discussed.  Allowing an immediate comparison of
several options with identical physics performance allows the
determination of the most cost efficient solution.
Emphasize is given on the possible
application of solid light concentrators with their typically about two times better concentration allowing the use of small photo sensors
such as Geiger-mode avalanche photo diodes. This is discussed in more details
in the context of possible design options for the Cherenkov Telescope Array.
In particular, a solution for a 60\,mm\(^2\) photo sensor with hollow cone is
compared to a 36\,mm\(^2\) with solid cone.

\end{abstract}


\end{frontmatter}

\section{Introduction}
The next generation Cherenkov Telescope Array (CTA) will be constituted
of three types of imaging atmospheric Cherenkov telescopes: small-,
medium- and large-sized telescopes. The array will be dedicated to the
observation of the high energy gamma-ray sky with unprecedented
sensitivity~\cite{cta-tdr} over a broad range of energies 
(0.01\,TeV\,\(\lesssim{}E_\gamma\lesssim\)\,100\,TeV). This instrument
will enable astronomers and astro-particle physicists to refine models
of gamma-ray sources and underlying non-thermal mechanism at work, to
study the origin and the composition of the cosmic rays up to the knee
region, question the nature of the dark matter, etc.

These kind of telescopes detect Cherenkov light emitted along
developing showers in the atmosphere nearly uniformly illuminating the
ground over an area of about 50'000\,m\(^2\). Shower maxima of a
gamma-ray induced electromagnetic shower occur at altitudes comprised
between \(\sim\)7\,km and \(\sim\)12\,km, for a primary energy ranging
from \(\sim\)100\,GeV to \(\sim\)100\,TeV.


The photon density on the ground depends on the energy of the gamma
ray, its incidence direction and the distance from the resulting shower
axis. The displacement of the image centroid in the telescope camera is
a function of the angular distance of the shower core, defined by
shower impact distance and altitude of the shower and is typically
about 1.5\textdegree{} for an energy of 1\,TeV and an impact parameter
of 150\,m, and 3\textdegree{} to 3.5\textdegree{} for energies around
100\,TeV with an impact parameter of 300\,m. As very high energy
showers penetrate deeply in the atmosphere and generate a large amount
of light, they can be observed at a relatively large distance from the
main light pool already with relatively small reflectors translating
into the necessity of a deployment of telescopes with large
field-of-view for the exploration of very high energy showers. In
general, the required telescope field-of view in order to record
contained shower images ranges between 3\textdegree{} and
10\textdegree{} depending on the energy range of interest.

A promising design for wide field observations is the Davies-Cotton
telescope~\cite{mirzoyan}. Its advantage  is a point-spread function
enabling a larger field-of-view than a parabolic design. The
Schwarzschild-Coud\'e design (\eg~\cite{Vassiliev}) is not considered
further, as it presents technical and cost challenges compared to 
a conservative CTA proposal using a prime optics design (see also
the comments in section~\ref{sct:winstonCone}).

It is demonstrated that once the field-of-view of the camera pixels and
of the whole instrument together with the photo-sensor technology is
fixed by means of physics arguments, only one parameter is left free.
This parameter can be fixed as well, relating the worst optical
resolution in the field-of-view, \ie at the edge of the field-of-view,
with the size of a pixel. Having a reasonable cost model at hand, even
the most cost-efficient single telescope or telescope design operating
in an array can be derived.

Knowing the anticipated angular size of a pixel from physics
constraints, a requirement on the point-spread function of the
telescope can be determined. If the point-spread function is identified
by the root-mean-square of the light distribution, requiring a
pixel diameter four times the root-mean-square ensures that most of
the light from a point source at infinity is concentrated in a single
pixel at the edge of the field-of-view.\footnote{In the case of a
normal 2D-Gaussian distribution this would correspond to more than 86\%
of the light distribution.} If simulations show that such a containment
is not necessary in terms of angular resolution of the shower origin or
background suppression, also a smaller value like twice the
root-mean-square can be considered, \cf~\cite{Aharonian1995}.
If this requirement is not optimized or not met, either the physics outcome
is worsened or the camera has more pixels than necessary and will not be 
cost-efficient.

The CTA array layout will consist of large size telescopes
(LST, primary mirror diameter \(\sim\)24\,m) in the center, and
successively surrounded with an increasing number of medium-sized (MST,
\(\sim\)12\,m) and small-sized (SST, \(\sim\)6\,m) telescopes in order
to instrument a ground surface area comprised between 4\,km\(^2\) and
10\,km\(^2\). Telescope spacings, sizes and field-of-views 
reflect the energy range to be explored by a certain type of
telescope. The LST sub-array will be primarily focusing on the
observation of the high energy gamma-ray sky with great precision below
about 100\,GeV, the inter-telescope spacing will be small, about 60\,m
and the single telescope field-of-view limited to about
3\textdegree\,--\,4\textdegree. The SST sub-array will conversely be
optimized for multi-TeV observations, up to around 100\,TeV allowing
for large spacing of up to 300\,m\,--\,500\,m with relatively modest
reflector size. The camera pixel field-of-view of these different
telescope types will span between about
0.08\textdegree\,--\,0.1\textdegree{} (LST) up to
0.2\textdegree\,--\,0.3\textdegree{} (SST), linked to the different
shower intensity with its intrinsic fluctuations and the concurrent
necessity of keeping the number of camera pixels reasonably low.

Consequently, this study will focus on reflector diameters in the range
of a meter to about 30\,m and on pixel field-of-views in the range between
\(\sim\)0.08\textdegree{} and \(\sim\)0.3\textdegree.
and is extended to off-axis angles of the incoming light up to more than
5\textdegree{}.

In the following, first a description is derived for the relations
between the existing design parameters of a Cherenkov telescopes, \eg\
focal length and reflector diameter. Then, by including a
semi-analytical treatment for the optical quality of a generalized
Davies-Cotton reflector, this description becomes applicable  for the
design of real telescopes. At the end, the influence of the variation
of the parameters on the optimized design is discussed.

\section{Optimization}

The light collection is one of the most important parameter
of a Cherenkov telescope and can be improved by, \eg, an increase of
the photo-detection efficiency of the photo-sensors or a rescaling of
the system, \ie, a corresponding rescaling of the reflective area and
photo-sensor size. As current technology (photo-multiplier tubes,
hybrid photo-detectors or silicon photo-multipliers) only offer a limited
choice of photo-sensor sizes, the Cherenkov telescope design
parameter phase space is reduced. Given a particular photo-sensor type, the
addition of a light concentrator in front of the photo-sensor is the
only way to increase the light collection efficiency. 

In the following, it will be shown that the relationship between the
characteristic parameters of an optimal telescope design (pixel
field-of-view and linear size, reflector diameter and focal distance)
is fully constrained, once the technological requirement (photo-sensor
size and the light-guide material) and the physics requirements (the
pixel angular size and the field-of-view) are frozen. This result is
obtained by imposing restrictions on the value of the
point-spread function at the edge of the field-of-view in order to keep
adequate image quality and thus analysis potential.

\subsection{Light concentrators}\label{sct:winstonCone}
The theorem of Liouville states that the maximum concentration
theoretically achievable is defined by maintaining the phase space,
\ie the product of solid angle, defined by the incoming light rays
directions, light ray momentum squared and the surface area crossed by
the light rays. The theorem of Liouville is applicable to the case of a
Winston cone with entrance area \(A\) placed in front of a photo-detector
of area \(A'\), corresponding to the exit area of the cone. \(A\) is
defined provided \(A'\), the solid angle \(\Omega\) defined by the light
rays entering the cone and the solid angle \(\Omega'\) defined by the
light rays leaving the cone.

Winston has shown in~\cite{winston} that the maximum concentration factor
for a rotationally symmetric light concentrator is
\[C_{max} = \frac{A}{A'} = \left(\frac{n'}{n}\right)^2\cdot\frac{1}{\sin^2\phi}\,, \label{eqn:concentration}\]
where \(n\) and \(n'\) denote the refractive index of the media in
front of and inside the optical system with \(n\)\,\(\approx\)\,1 in
air. \(\phi\)~corresponds to the maximum angle at which a light ray
enters the system related to the solid angle \(\Omega\). 

If the system is not axisymmetric or the angular acceptance of the
photo detector is smaller than \(\Omega'= 2\pi\) (as assumed in
Eq.~\ref{eqn:concentration}), \(C_{\mathrm{max}}\) has to be adapted
accordingly.

Besides the increase of the light collection area, the use of
cones enables a partial screening of the night-sky light pollution
corresponding to \(\phi\) being larger than the angle of light rays 
coming from the edge of the reflector.

Simulations~\cite{braun} and recent concentration efficiency
measurements~\cite{proceeding-IBraun} of solid cones
(\(n'\)\,\(\approx\)\,1.4) designed for the FACT
camera~\cite{LatestFACTref} demonstrated that their shape is nearly
ideal, that the concentration factor reaches a value close to
\(C_{\mathrm{max}}\) and the geometric loss is only of the order of a
few percent excluding absorption loss\footnote{The cones designed for
the FACT camera where designed on the assumption of a later extension
of the mirror area, \ie, they were designed for a larger reflector
diameter than the currently installed one.}. In the case of hollow
cones, Fresnel reflection losses have to be considered at the surface
of the photo-sensor. If the camera is sealed with a protective window,
which is usually the case, also losses at the window surface need
to be taken into account. By choosing a material for the cones and
the protective window with a similar refractive index than the 
material of the photo-sensor light entrance, these losses can be
omitted. Combined with the almost perfect reflectivity
of solid cones due to total reflection (limited only by the surface
roughness), solid cone usually outperform hollow cone.

The concentration factor achieved with the Winston cones is
fundamentally similar to the size reduction of the focal plane in a
Schwarzschild-Coud\'e design and linked to the conservation of the
space-momentum phase space according to the Liouville theorem: the
conversion of the spatial into momentum phase space, by means of 
Winston cones or secondary optics. While cones reduce the acceptance
of the incoming light rays from a large area at the entrance to a large
angular acceptance and small area at the cone exit, the secondary
mirror optics leads to a similar spatial compression and angular
widening of the light rays at the photo-sensor and thus a reduced
plate-scale. While cones are non-imaging devices, the secondary optics is
imaging. Hence, in the Schwarzschild-Coud\'e design, it is possible to
attain excellent optical resolution, in terms of Cherenkov telescope
requirements, with a field-of-view as large as 15\textdegree. Both
designs enable compression of the photo sensitive area by factors larger
than ten w.r.t.\ their primary optics design.

\subsection{Connection to the optical system}

In the case of a Cherenkov telescope, the light entering the cone comes
from a reflector visible under a maximum angle
\(\phi\)~\footnote{\(\phi\) defined from the focal plane center}. 
The opening angle of the light at the entry of the cone
is therefore well defined by the properties of the optical system, \ie
by the diameter of the reflector \(D\) and the focal length
\(F\), \(f=F/D\).
\[\tan\phi = \frac{1}{2f} \label{eqn:ConeOpeningAngle}\]

Combining this with Eq.~\ref{eqn:concentration} yields
\[\frac{A}{A'} = \left(\frac{n'}{n}\right)^2\cdot\left(4f^2+1\right)\,. \label{eqn:BaseFormula}\]
For instance, taking the FACT values, \(n'=1.4\) and \(f=1.4\), we
obtain the theoretically maximal achievable  concentration factor 
\(C_{\mathrm{max}}=17.3\), \ie the linear size of the entrance
area can ideally be larger than four times the linear photo-sensor
size.

\subsection{The optical system}

In addition, the optical system defines the zoom factor or
plate-scale, \ie the field-of-view corresponding to a physical
area in the focal plane. 
The correspondence between the angular size \(\vartheta\) and the
linear size \(\delta\) on the focal plane is
\[ \tan\frac{\vartheta}{2} = \frac{\delta}{2F} \label{eqn:PlateScale}\]
or in the limit of small \(\vartheta\),
\[ \delta \approx \vartheta F\,. \]

Cameras in Cherenkov telescopes are pixelized due to the use of photo-detectors. To increase the light collection efficiency further, and to
maintain symmetry, these pixels are usually aligned on a hexagonal grid,
\ie in closed package geometry. In recent years, MAGIC has
exploited the photon arrival time extracted from the measured pulse and
demonstrated significant improvements in the
sensitivity~\cite{magic-time}.

The technique, taking into account the change of the arrival time
between neighboring pixels, performs best, if all neighbors are at an
identical distance from the central pixel. Consequently, the ideal
shape of a pixel is hexagonal.

The distance \(\delta\) on the camera surface is the
distance between two parallel sides of a hexagon, its area is
\[ A = \frac{\sqrt{3}}{2}\,\delta^2\,. \label{eqp:AreaHexagon}\]
Combining with Eq.~\ref{eqn:BaseFormula}, the plate-scale formula
Eq.~\ref{eqn:PlateScale} and Eq. \ref{eqp:AreaHexagon},

\[\frac{\sqrt{3}}{2}\left(\frac{n}{n'}\right)^2\frac{\tan^2{(\vartheta/2)}}{A'} = \frac{1}{D^2}+\frac{1}{4F^2} \label{eqn:DerivedFormula}\]
is obtained, which translates the close relationship between the pixel field-of-view
\(\vartheta\), the focal length \(F\) and the reflector diameter \(D\),
once the technological parameters fixed: photo detector size \(A'\) and
light concentrator material \(n'\).

Defining a constant related to these properties 
\[ k(n,n',A') =  \frac{\sqrt{3}}{2\cdot A'}\cdot\left(\frac{n}{n'}\right)^2 \label{eqn:SystemConstant}\]
and rewriting Eq.~\ref{eqn:DerivedFormula}
(\(\tan\vartheta\approx\vartheta\) for \(\vartheta\ll\)1\textdegree) as
\[F = \frac{1}{\sqrt{k\,\vartheta^2-\left({2/D}\right)^2}}, \label{eqn:FocalLength}\]
it is immediately apparent that the focal length \(F\) of the system is
a direct consequence of the pixel field-of-view and the reflector
diameter, if the properties of the photon detector and the material
of the cones are known.
For typical Cherenkov telescopes, \(F/D\) is between unity and two.
Below unity the resolution
becomes too coarse and above two, not only the number of pixels and hence
the price of a camera becomes too high, but also the camera holding
structure becomes mechanically complex and hence disproportionally
expensive. This constraint on \(F/D\) applied to Eq.~\ref{eqn:FocalLength} yields
\[4.25<k\,\vartheta^2\,D^2<5\,.\]

Precisely the choice of \(F\) and \(D\) defines the optical quality of
a mirror system. At the same time, the size of a single pixel defines a
natural constraint on the optical quality of a system, \ie, \(F/D\)
should be chosen such that the point-spread function at the edge of
the camera is within a limit well defined by the pixel's
field-of-view.

\subsection{Optical quality}
The light collection area is important for a Cherenkov telescope
and typical reflector sizes range from a few to ten or twenty meters.
However, with the current technology, it is not possible to produce
large mirrors with the requested quality at a reasonable cost.
Furthermore, optical systems compiled from a single mirror suffer large
aberration effects at large off-axis angles, while a wide field-of-view
is necessary for the observation of multi-TeV showers up to large
impact parameters, as well as for extended sources.
Therefore, segmented mirrors are in use. The layout providing the
best optical quality for segments of identical focal length is the
so-called Davies-Cotton layout~\cite{dc-design}, where the single
spherical mirrors are located on a sphere with radius \(F\) and
focused to a point at \(2F\).

The relevant quantity which influences the on-axis and off-axis optical
quality is the focal ratio \(F/D\). The optical quality
improves with larger values. This scale invariance statement is true
only as long as the optical quality of a single mirror can be neglected
against the optical quality of the whole system, which is generally the
case at the edge of the camera.


To be able to constrain the optical point-spread function,
a relation between the tessellation, the focal ratio and
the resulting point-spread function is needed for a given maximum
inclination angle of the light, \ie at the edge of the field-of-view 
of the camera.
\begin{figure*}[htb]
\centering
 \includegraphics*[width=0.48\textwidth,angle=0,clip]{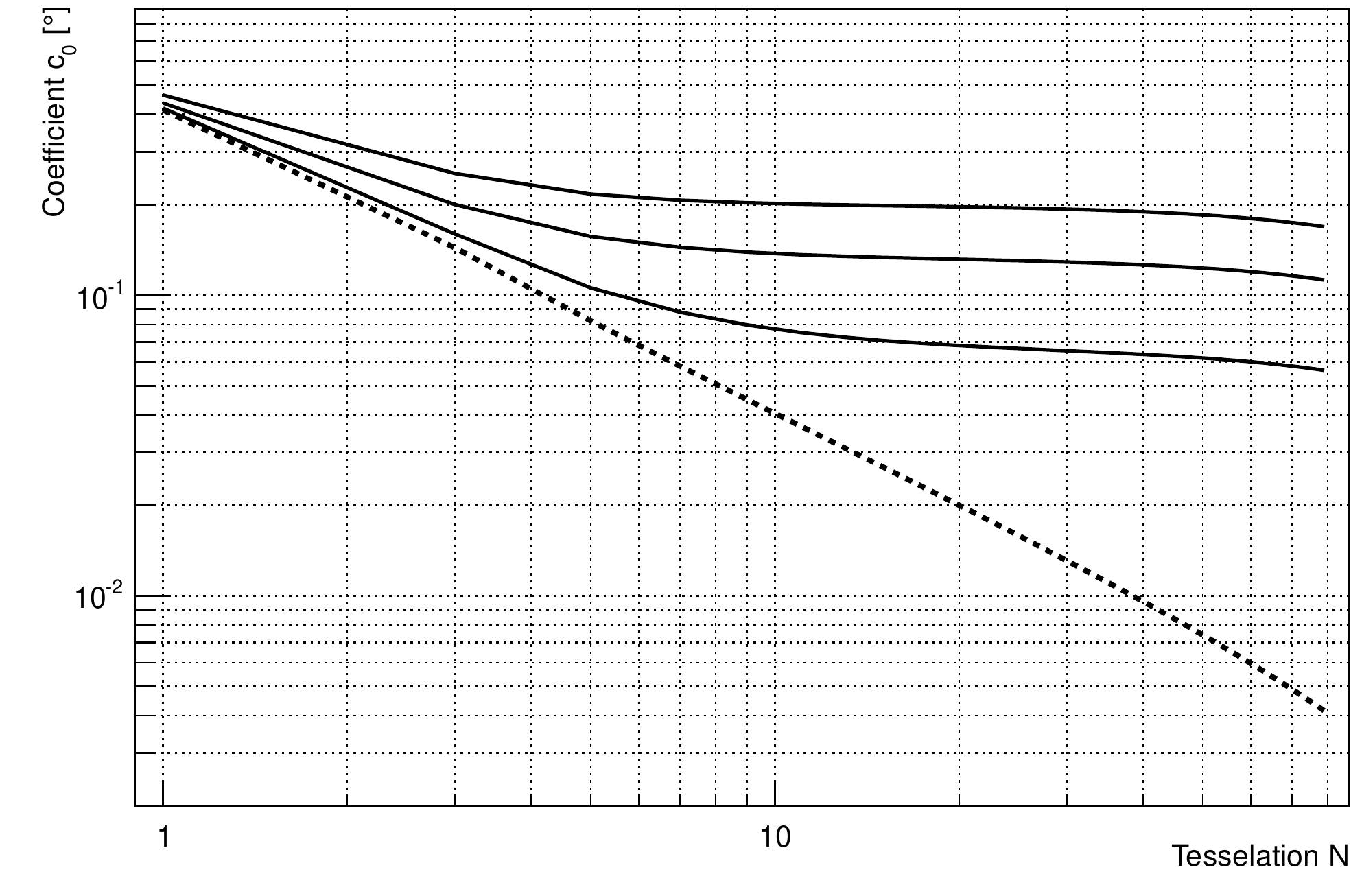}
 \includegraphics*[width=0.48\textwidth,angle=0,clip]{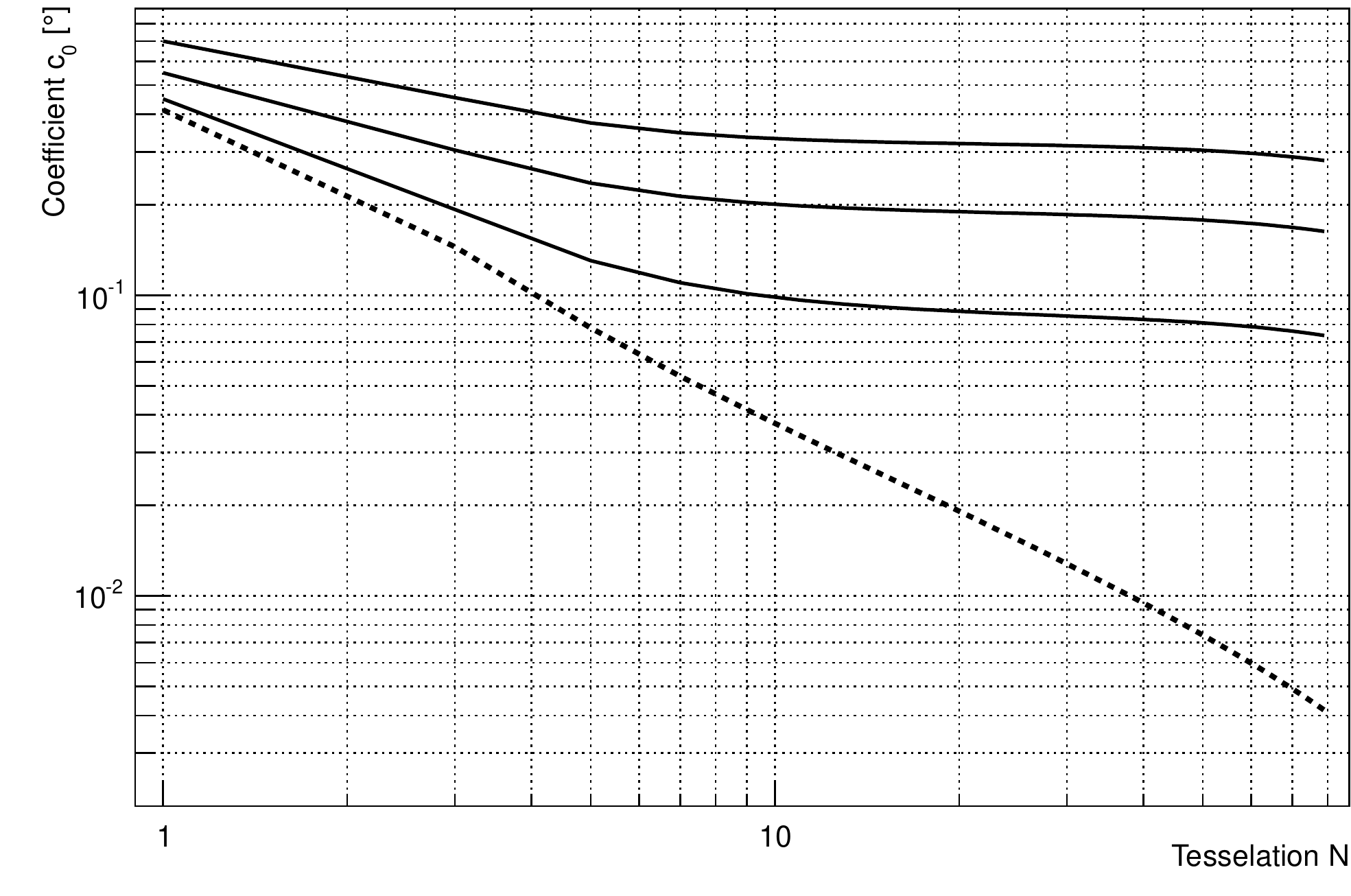}
 \includegraphics*[width=0.48\textwidth,angle=0,clip]{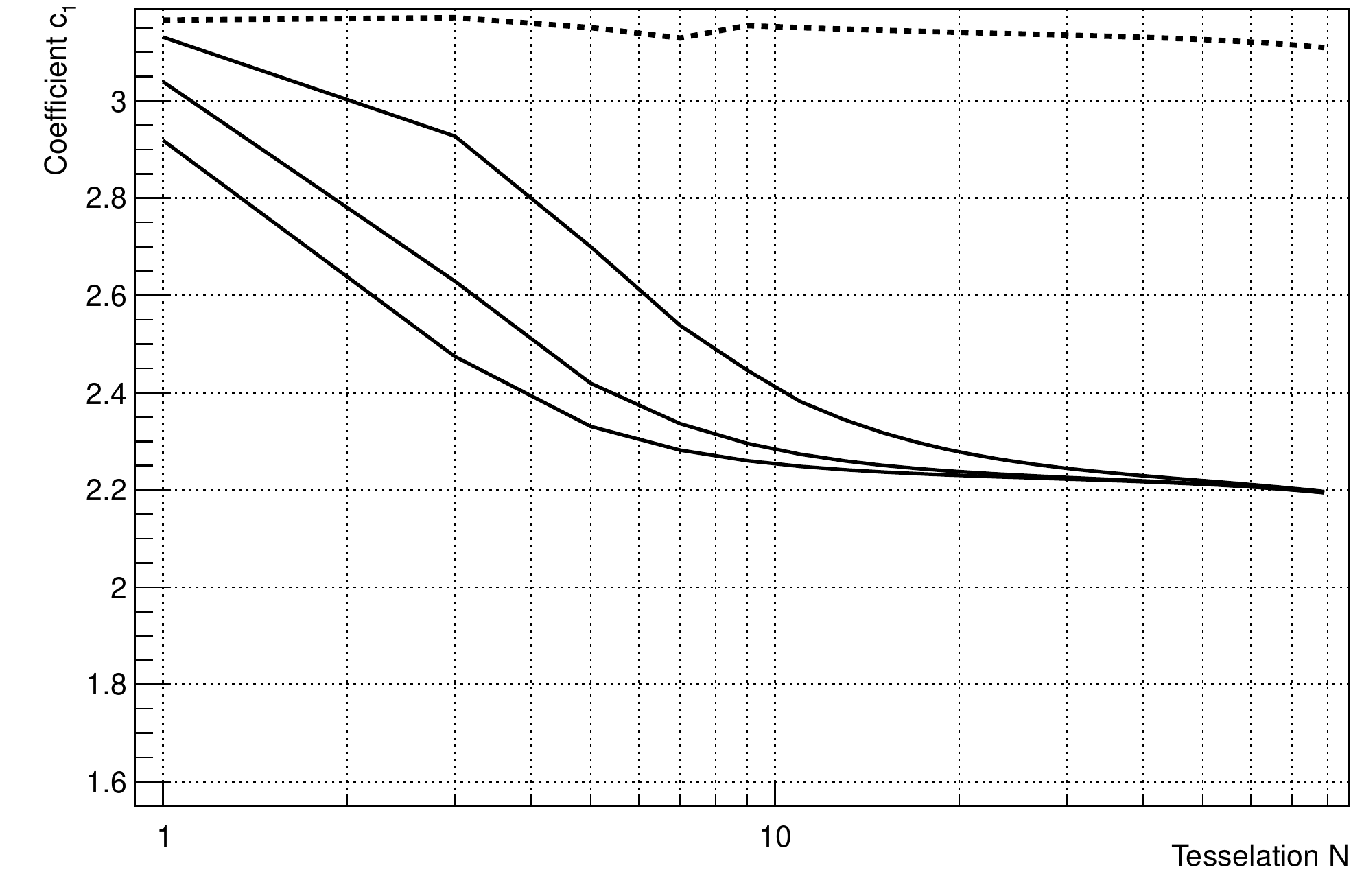}
 \includegraphics*[width=0.48\textwidth,angle=0,clip]{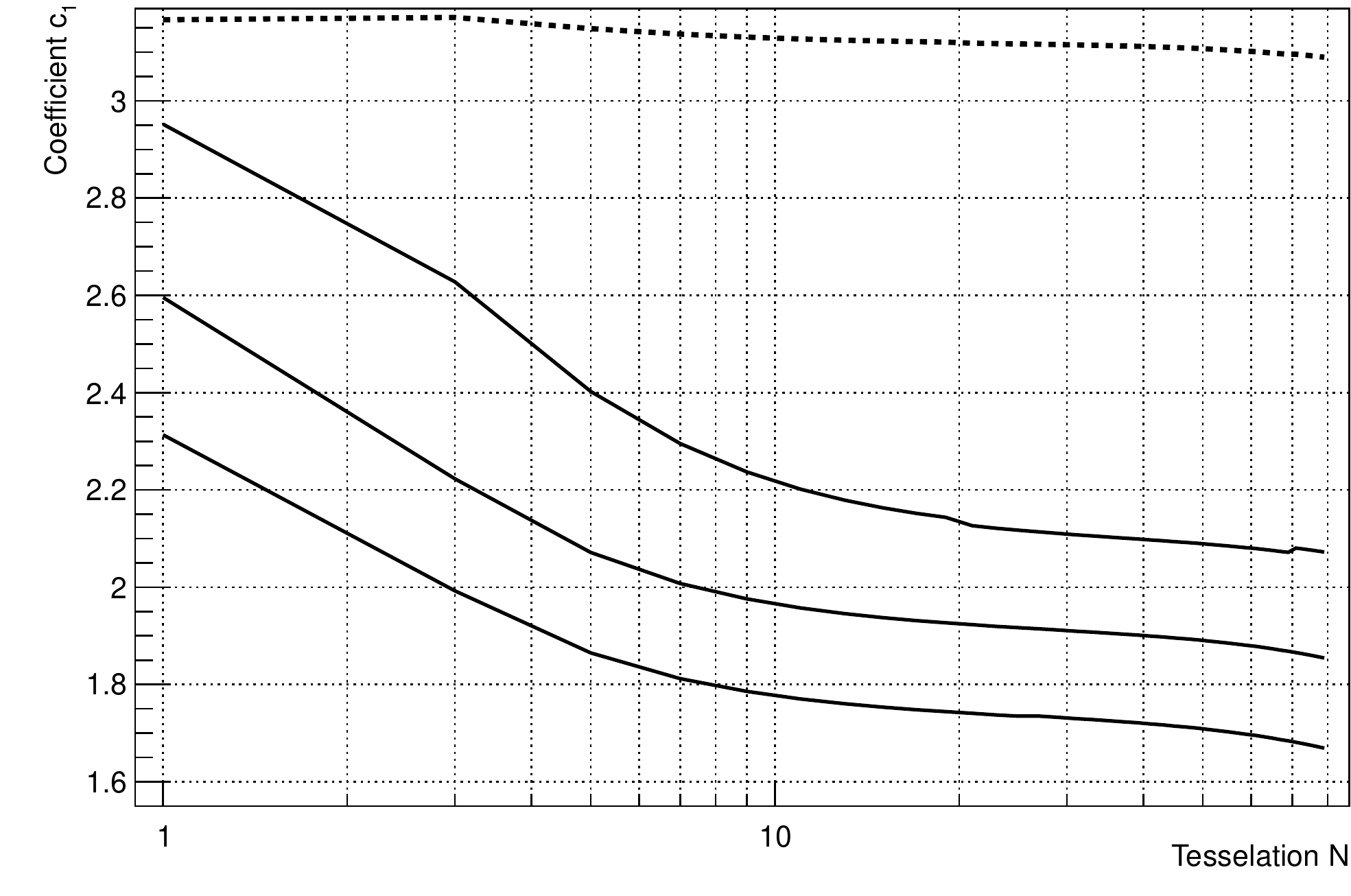}
\caption{Coefficients derived from the work in the Appendix to convert
the focal ratio \(f\) into point-spread function
as defined in Eq.~\ref{eqn:PointSpreadFunction}.
Sagittal component (left) and the tangential component (right). The dashed
line denotes on-axis rays \(\alpha\)=0\textdegree, the solid lines (from
the dashed line outwards) rays at 2\textdegree, 4\textdegree{} and
6\textdegree{} off-axis.}
\label{fig:coefficients}
\end{figure*}

In~\ref{appendix:A}, a formalism describing the point-spread function
of an ideal Davies-Cotton reflector, \ie a reflector with infinite
tessellation is presented. The point-spread function is described
by the root-mean-square of the light distribution. With the help of
ray-tracing simulation, a reasonably good description of a real
tessellated Davies-Cotton reflector is derived from this analytical
approach in~\ref{appendix:B}. Including the correction factor which
describes the deviation of the analytical approach from the
simulations, a good description of the point-spread function is
obtained.
It is shown that the point-spread function \(P\) of an ideal
Davies-Cotton can be expanded into a polynomial in
\(\alpha^i\cdot{}f^{-j}\). Given the incident angle \(\alpha\ll 1\) of
the incoming ray and \(f\) in the range between one and two, the
polynomial is hypothesized into the more simpler form
\[ P(f,\alpha,N) = c_0(\alpha,N)\cdot{}f^{-c_1(\alpha,N)} \label{eqn:PointSpreadFunction}\]
with coefficients \(c_0\) and \(c_1\) and \(N\) the tessellation number as
described in~\ref{appendix:B}. This parametrization is found to match
the ray-tracing simulation without loss of precision.
The coefficient \(c_0\) can directly be deduced as the result of
Eqs.~\ref{eq:realDC} at \(f=1\) and \(c_1\) is derived by
a fit. An example for the coefficients \(c_0\) and \(c_1\) for
selected \(\alpha\) is shown in Fig.~\ref{fig:coefficients}.

\subsection{Result}


As discussed in the introduction, it is required that the 
point-spread function is small compared to the pixel field-of-view
at the edge of the field-of-view, so that the light of a point source
is well contained in one pixel. 
Defining the ratio \(r\) between both, this requirement can be expressed as
\[ \vartheta = r\cdot P\,.\label{eqn:PSF_relation}\]
Combined with Eq.~\ref{eqn:PointSpreadFunction}, the focal ratio \(f\) can now be expressed as
\[ f = \left(\frac{\vartheta}{r\cdot c_0}\right)^{-\frac{1}{c_1}}\,.\label{eqn:FocalLength2}\]
Including this in Eq.~\ref{eqn:FocalLength} yields
\[ D^2 = \frac{4\,g}{k}\ \vartheta^{-2}\label{eqn:Result}\]
with a correction factor \(g\) defined as
\[g=1+\frac{1}{4}\,\left(\frac{\vartheta}{r\cdot c_0}\right)^{\frac{2}{c_1}}\,.\label{eqn:f}\]
The absolute focal length \(F\) can now be calculated using
Eq.~\ref{eqn:FocalLength}.

\begin{figure*}[htb]
\centering
 \includegraphics*[width=0.48\textwidth,angle=0,clip]{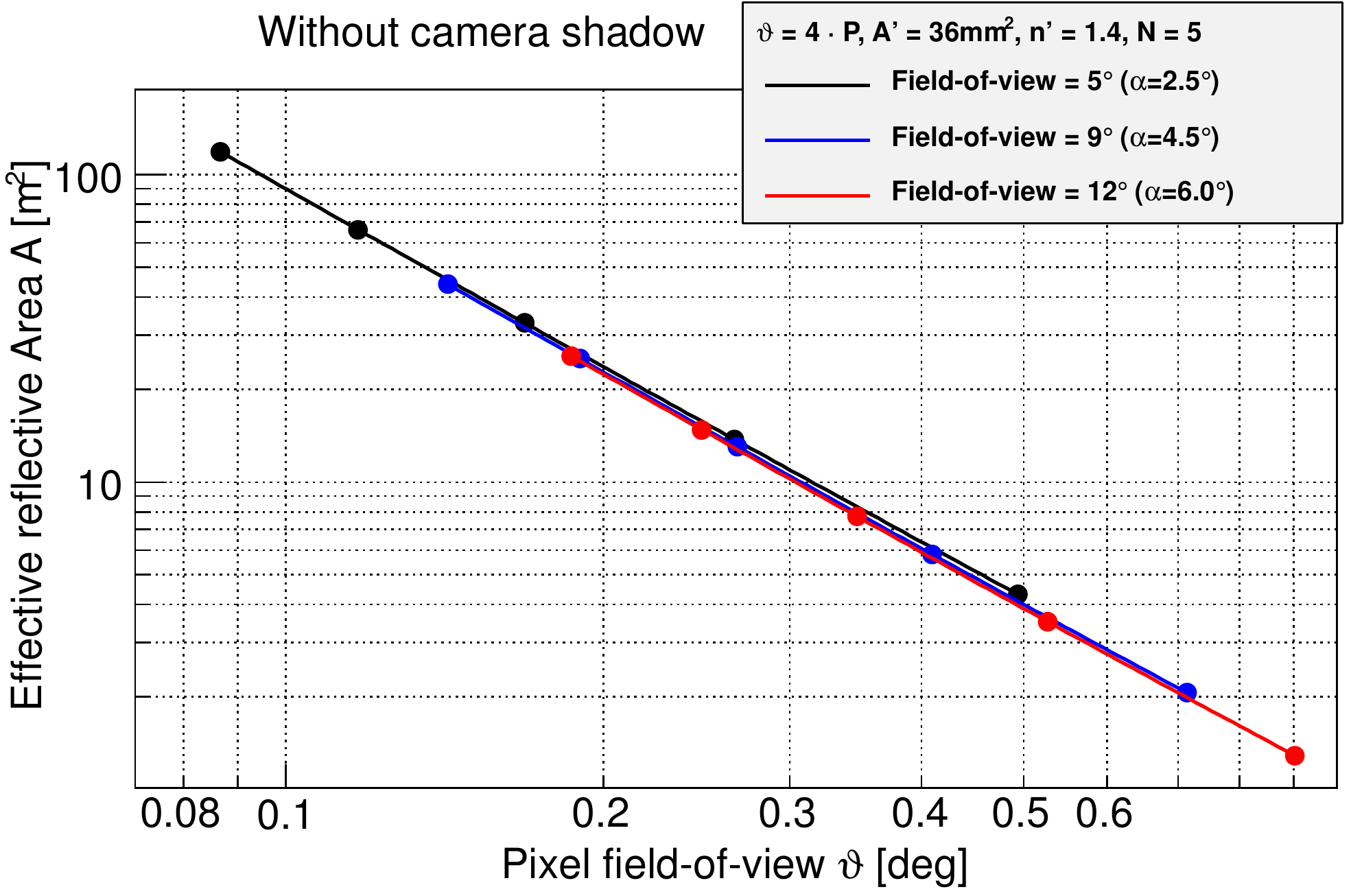}
 \hfill
 \includegraphics*[width=0.48\textwidth,angle=0,clip]{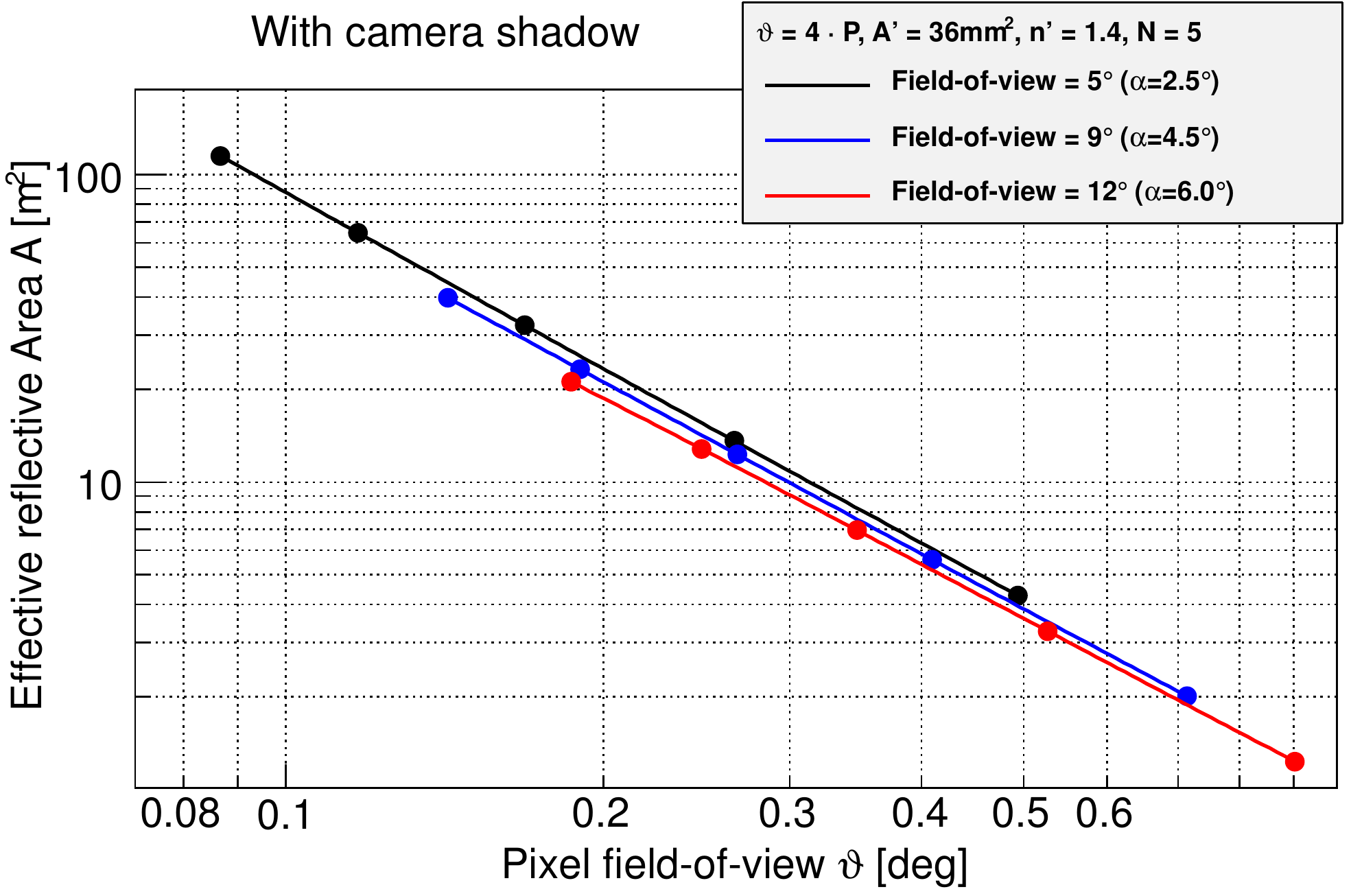}
\caption{Reflective area (left) and the effective reflective
area (right), \ie including the camera shadow, for the standard setup as discussed
in section~\ref{Discussion}. The result is shown in the range between
\(F/D\)\,=\,[1, 2]. The dots denote intermediate results at \(F/D\)\,=\,\{1.25, 1.5, 1.75\}.
For convenience the result is also shown for a camera field-of-view of 5\textdegree{}
and 12\textdegree. The legend gives the corresponding pixel field-of-view \(\vartheta\) with its relation to the optical point-spread function \(P\) at the edge of
the camera, the photo sensitive area \(A'\) of the photo sensor,
the refractive index \(n\) of the light collector and the tessellation of the reflector expressed as the number \(N\) of individual mirrors on its diagonal.\label{fig:ResultA}}
\end{figure*}

To deduce the effective reflective area from Eq.~\ref{eqn:Result}, the 
shadow of the camera on the reflector has to be taken into account. To
calculate the fraction of the reflector shadowed, the ratio of their
areas is calculated. Conversion of \(\alpha\) from an angle to a length
yields approximately \(R=F\cdot\alpha\) for small values of \(\alpha\)
(\(\ll\)10\textdegree). 

Expressing the focal length by Eq.~\ref{eqn:FocalLength2} the fraction
of the camera shadow on the reflector is derived.

\[\frac{A_{cam}}{A_{ref}} = 
\frac{\pi F^2\alpha^2}{\pi\left(\frac{D}{2}\right)^2} = 
\left(2\alpha f\right)^{2}=\frac{\alpha^2}{g-1}\]

If a real camera housing is significantly larger than the photo sensitive
area itself, a correction factor should be included.

Now the effective light collection area of the optical system can be deduced as
\[A_{eff} = \frac{\pi g}{k}\left(1-\frac{\alpha^2}{g-1}\right)\vartheta^{-2}\,. \label{eqn:EffectiveArea} \]

If real setups should be compared, like \eg Davies-Cotton and
Schwarzschild-Coud\'e, also other sources of light-losses must be included,
such as geometrical efficiency of the cones (light-loss at the
edge of the mirror), total mirror reflectivity, cone transmission or
reflection losses, or photo detection efficiency.

\subsection{Discussion}\label{Discussion}

The relation given in Eq.~\ref{eqn:Result} includes several parameters
which are subject to change. For simplicity, a standard setup has been
defined to which altered setups are compared.

Silicon photo-detectors are a recent and very promising technology.
Therefore, a silicon photo-detector with a sensitive area
of 36\,mm\(^2\) is chosen as a benchmark device. Such devices
are already commercially available with acceptable properties.
Although their sensitive area is still rather small compared to
photo-multipliers, by increasing their light-collection with
solid light concentrators, their light-collection area becomes
reasonably large. Such light-concentrators still maintain
a reasonable weight and length in term of absorption.
Typical Plexiglas materials have a refractive indices of the order of
\(n'\)=1.4 and are used hereafter as a reference.



As the light-collection area of a telescopes scales directly with the
photo-sensitive area, the most obvious use of small photo sensors is
a small telescope sensitive mostly to high energetic showers. At high
energies, the collection area of a telescope array is of prime importance
due to rapidly decreasing fluxes.  Due to the bright
light-pool of high energy showers, telescopes with relatively small
reflectors can be operated with large spacing of, \eg, 400\,m or more,
\cf~\cite{Stamatescu}. Since such spacings demand a large camera field-of-view,
a field-of-view of 9\textdegree{} diameter is chosen as a reference.

A typical reflector for a Cherenkov telescope with Davies-Cotton
layout enables the manufacturing of a primary reflector tessellated
into spherical mirror of identical focal lengths. From the scaling with
the tessellation number as derived in \ref{eq:realDC}, it can be
concluded that a layout with only three mirrors on the diagonal (\(N\)=\,3) has
still a significantly worse optical quality than a reflector with five
mirrors on the diagonal. Although the point-spread function at the center of the camera
is clearly dominated by the mirror size, the relative influence almost
vanishes at higher off-axis angles. Since the solution with \(N\)=\,3
still shows a degradation of more than 10\%  compared to the solution
with \(N\)=\,5 even at the highest simulated off-axis angles, it is
discarded. On the other hand, a further increase of the tessellation
number (individual mirror size over primary reflector diameter) does not
significantly improve the optical quality. Consequently, choosing
\(N\)=\,5 is a good compromise and already close to the optimum
achievable. Comparable results were obtained in~\cite{Schliesser2005} although
using a third order approximation overestimating the optical quality.

It must be noted that the simulation does not take the point-spread function of the
individual mirrors nor any possible misalignment into account which must
be added quadratically to the result. However, for the solutions
discussed here this can be neglected, \cf~\cite{Bernlohr}. In general,
alignment errors can be kept minimal and individual mirrors can be machined
with a point-spread function small compared to the point-spread function at the
edge of the camera.

On average, all Davies-Cotton designs with a reasonable \(F/D\) have a
root-mean-square of the light distribution in the tangential direction about
two times larger than in the sagittal direction.

Ideally, the sagittal root-mean-square at the edge of the camera should
fit a fourth of the pixel's field-of-view. This ensures that in the sagittal
direction 95\% of the light is contained within one pixel diameter and
roughly 68\% in the tangential direction. However, since the point-spread
function is not Gaussian and has long tails in tangential direction exact
numbers for the light content might slightly differ.

For convenience, all following plots show dots for
\(F/D\)\,=\,\{1, 1.25, 1.5, 1.75, 2\}.

Fig.~\ref{fig:ResultA} shows the reflective area versus the pixel's
field-of-view for comparison in the standard case with and without
shadowing for different camera field-of-views.
Since the effect is comparably small and the mirror diameter is
more expressive, in the following all plots show the mirror diameter rather
than the reflective surface in the non-obstructed case.

\begin{figure*}[p]
\centering
 \includegraphics*[width=0.48\textwidth,angle=0,clip]{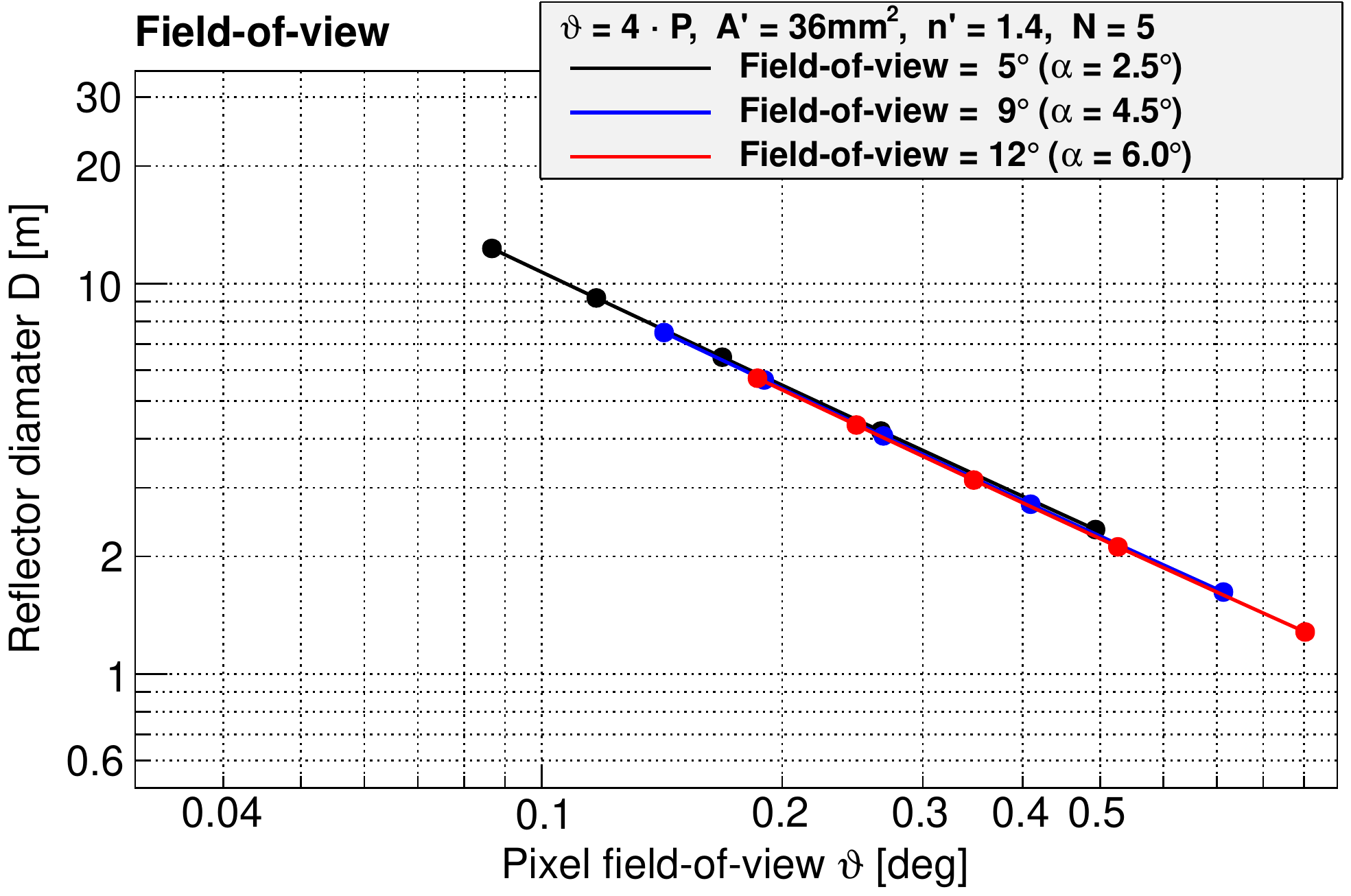}\\[3em]
 \includegraphics*[width=0.48\textwidth,angle=0,clip]{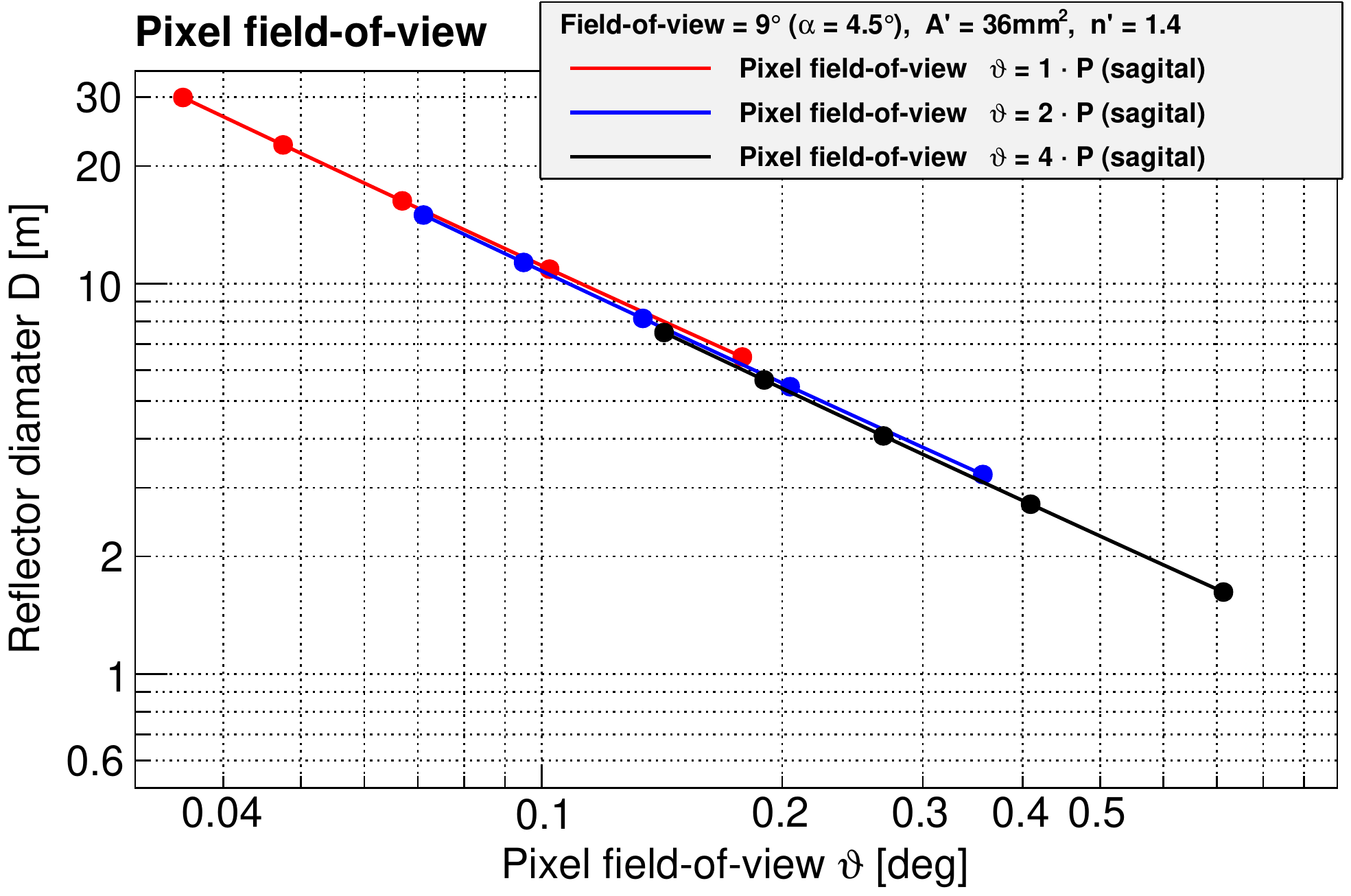}
 \hfill
 \includegraphics*[width=0.48\textwidth,angle=0,clip]{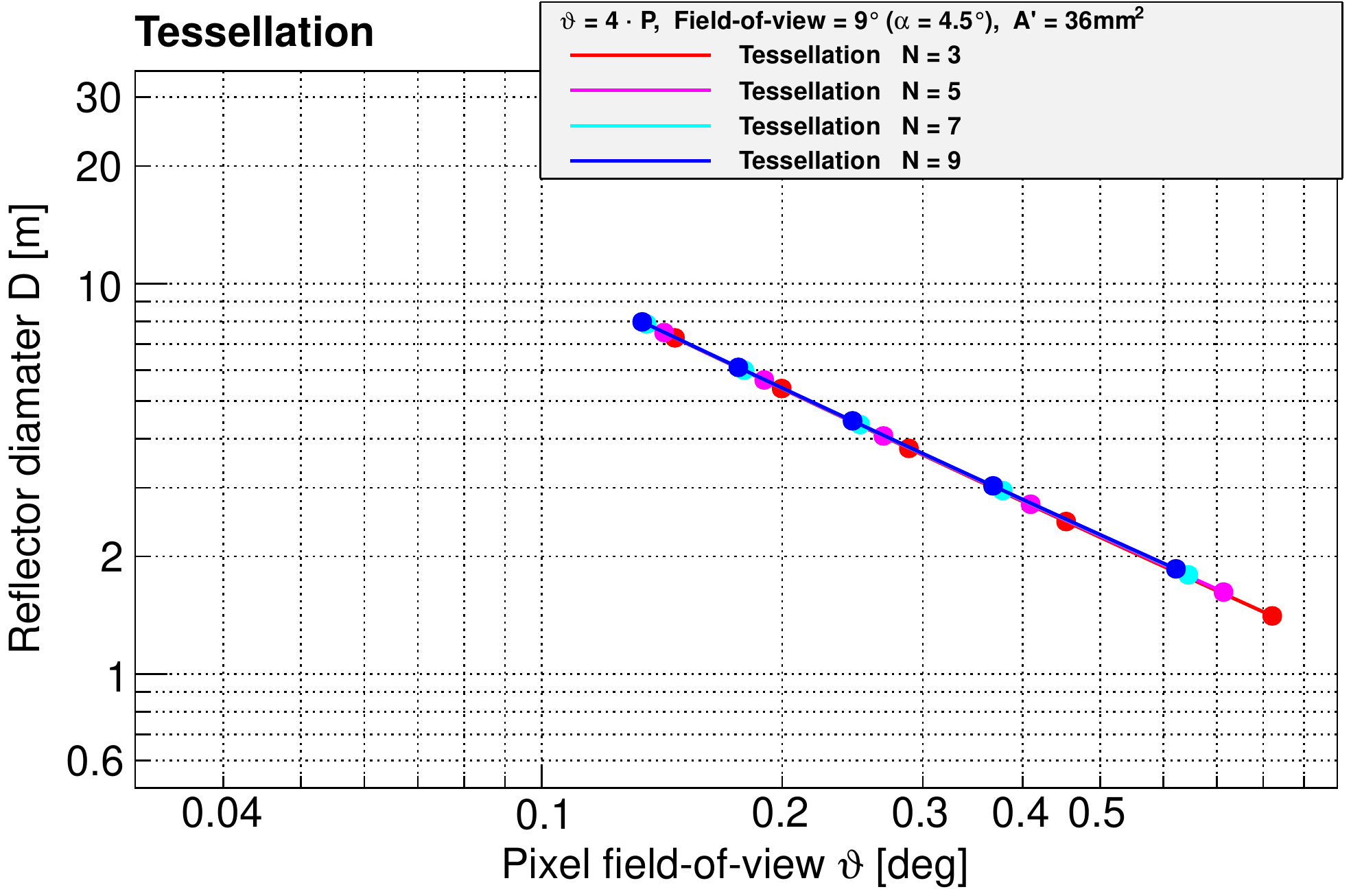}\\[3em]
 \includegraphics*[width=0.48\textwidth,angle=0,clip]{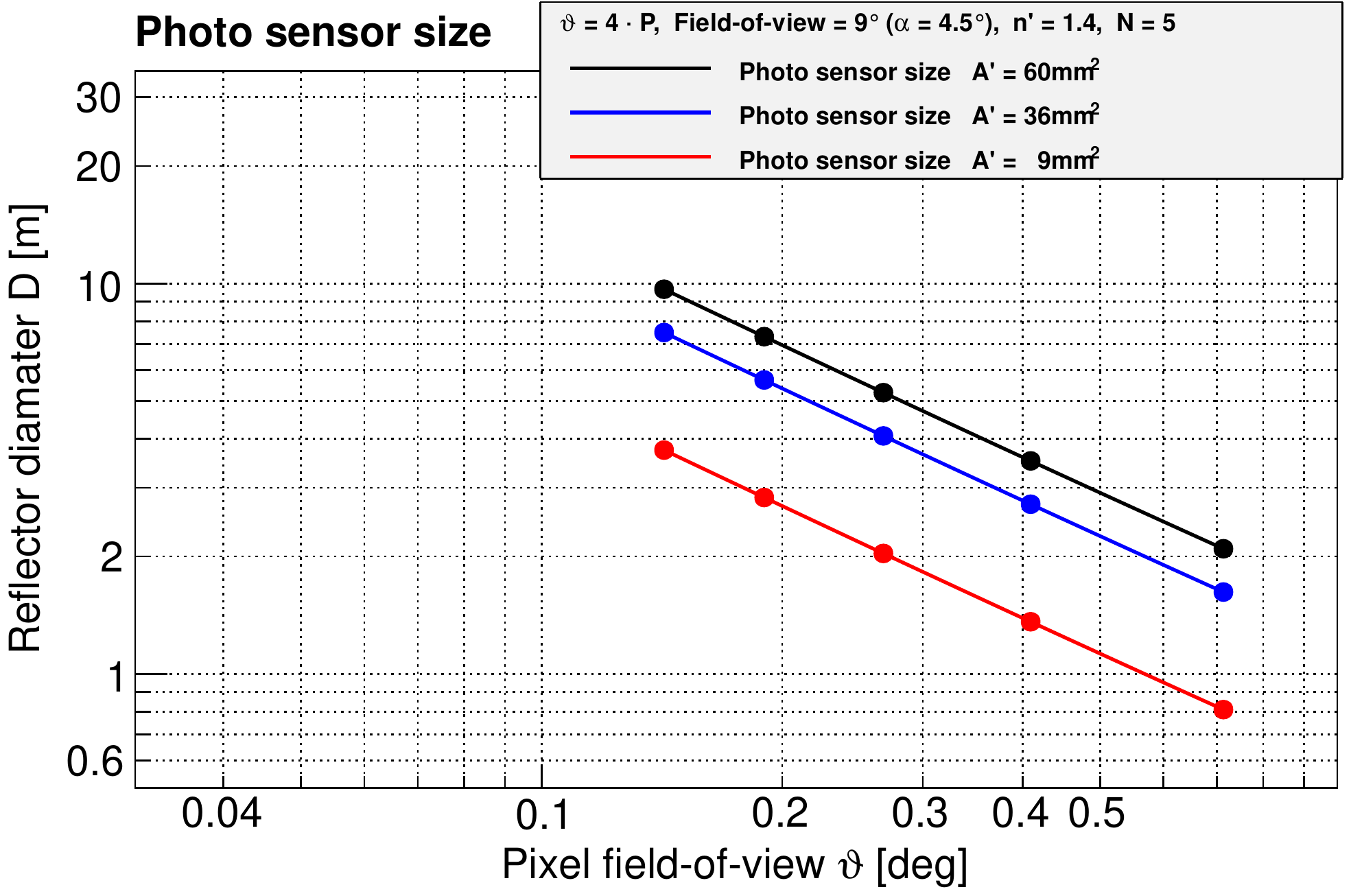}
 \hfill
 \includegraphics*[width=0.48\textwidth,angle=0,clip]{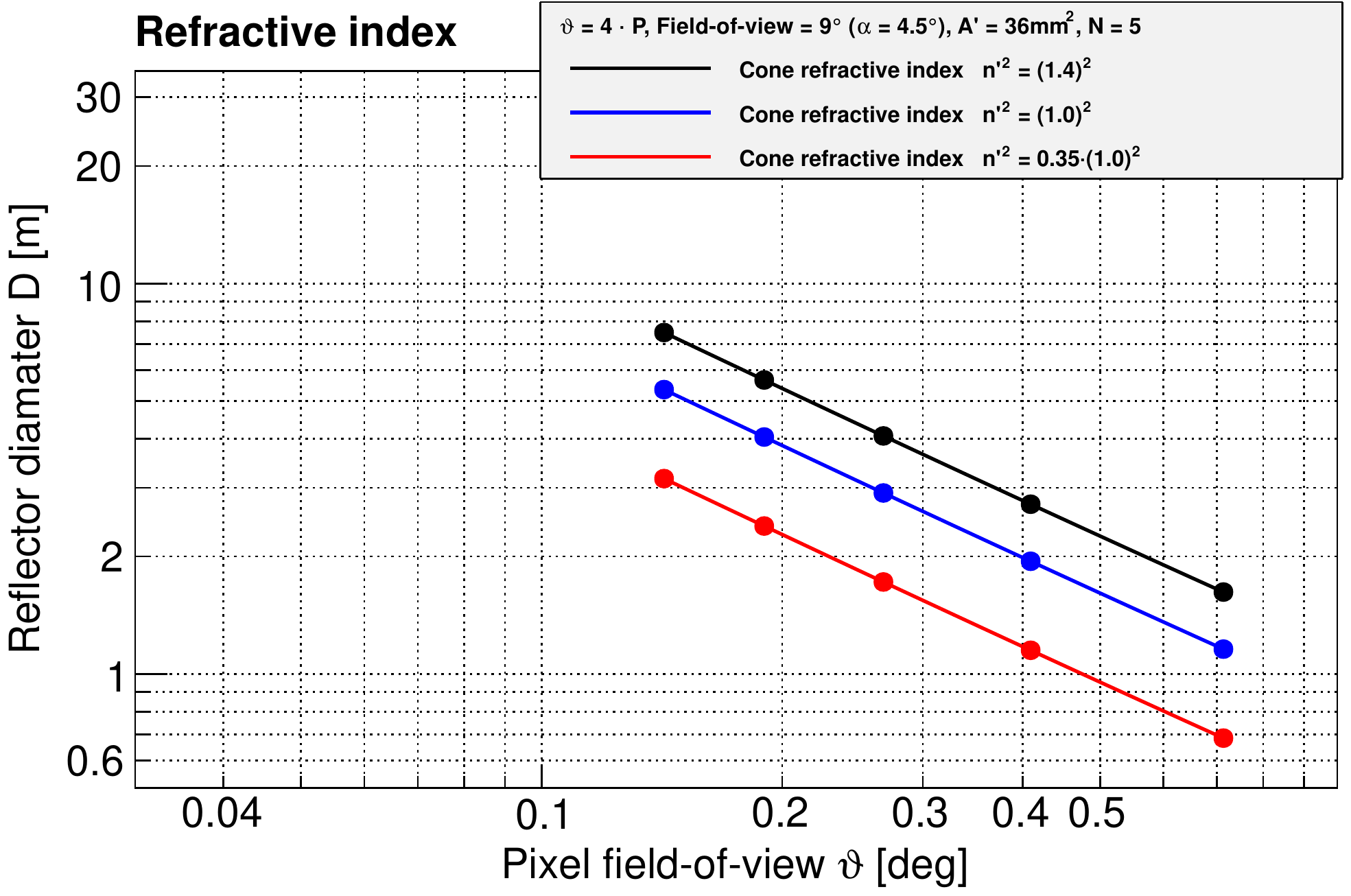}\\[3em]
\caption{Reflector diameter versus pixel's field-of-view for different 
setups. The results are shown in the range between \(F/D\)\,=\,[1,2]. The
dots denote intermediate results at \(F/D\)\,=\,\{1.25, 1.5, 1.75\}. For
convenience the result is also shown for a camera field-of-view of
5\textdegree{} and 12\textdegree. 
\label{fig:Comparison}}
\end{figure*}

The effects of changing different input parameters w.r.t.\ to the
previously described benchmark configuration are shown in
Fig.~\ref{fig:Comparison} and discussed below.

\vspace{-1ex}\paragraph{Changing the camera field-of-view (Fig.~\ref{fig:Comparison}, top plot)}

Changing the camera's field-of-view basically shifts the valid range
along the line, \ie the range corresponding to \(F/D\)\,=\,[1.0, 2.0]. 
That means that it is possible to build telescopes identical in optical
quality, pixel's field-of-view and mirror diameter, but different
field-of-view resulting simply in a different focal length of the
system. 
In short: Changing the field-of-view only changes the focal length.

\vspace{-1ex}\paragraph{Changing the optical quality (Fig.~\ref{fig:Comparison}, middle plots)}

A change in the requirement on the optical quality
\(r\) directly influences \(F/D\), and therefore also shifts the range
of reasonable \(F/D\) almost linearly in \(\vartheta\) (left plot). Changing the tessellation (right plot) is like changing the requirement on the
optical quality. While the difference in optical quality between
a Davies-Cotton layout with three mirrors on the diagonal and five mirrors
is still significant, all other layouts give identical results within
a few percent. 
In short: Any tessellation number \(\ge\)5 gives similar results.
Changing the requirement on the optical quality only changes the focal
length.

\vspace{-1ex}\paragraph{Changing the photo sensitive area (Fig.~\ref{fig:Comparison}, bottom left plot)}

Since the constant \(k\) is directly proportional to
the size \(A'\) of the photon detector, the mirror area is directly
proportional to the size of the photo sensor. 
If the size of the photon sensor is limited, a simple way to increase
the field-of-view of a single pixel is to sum the signal of several
photon counters to a single signal. To maintain a hexagonal, \ie
most symmetric layout, summing the signal of three, four or seven
photon sensors seems appropriate. 
In short: Assuming an optimized light-concentrator, the photo
sensor's physical size defines the scale of the system.

\vspace{-1ex}\paragraph{Changing the light concentrator (Fig.~\ref{fig:Comparison}, bottom right plot)}

Another way to increase the reflective area is an increase of the
refractive index of the light concentrator entering quadratically.
Using solid cones made from a Plexiglas material with a typical
refractive index in the order of 1.4 allows to increase the achievable
reflective area by a factor of two compared to hollow cones. Since the
length of a typical light concentrator for an exit of 1\,mm diameter is
in the order of 3\,mm\,--\,4\,mm, weight and light-attenuation, which
is dependent on the length of the material crossed, will define a
natural limit on the sensor size for which a solid cone is still
efficient. 
For comparison reasons not only solid (\(n'\)=1.4) cones but also
intentionally less efficient hollow cones (\(n\)=1.0) are shown.
Non-optimum hollow cones are typically used in current Cherenkov
telescopes, in which the sensitive area of standard photo-detectors
(PMTs) is not a limiting factor.
%
%
In short: Increasing the refractive index, quadratically increases
the reflective area of the system.


\begin{figure}[htb]
\centering
 \includegraphics*[width=0.48\textwidth,angle=0,clip]{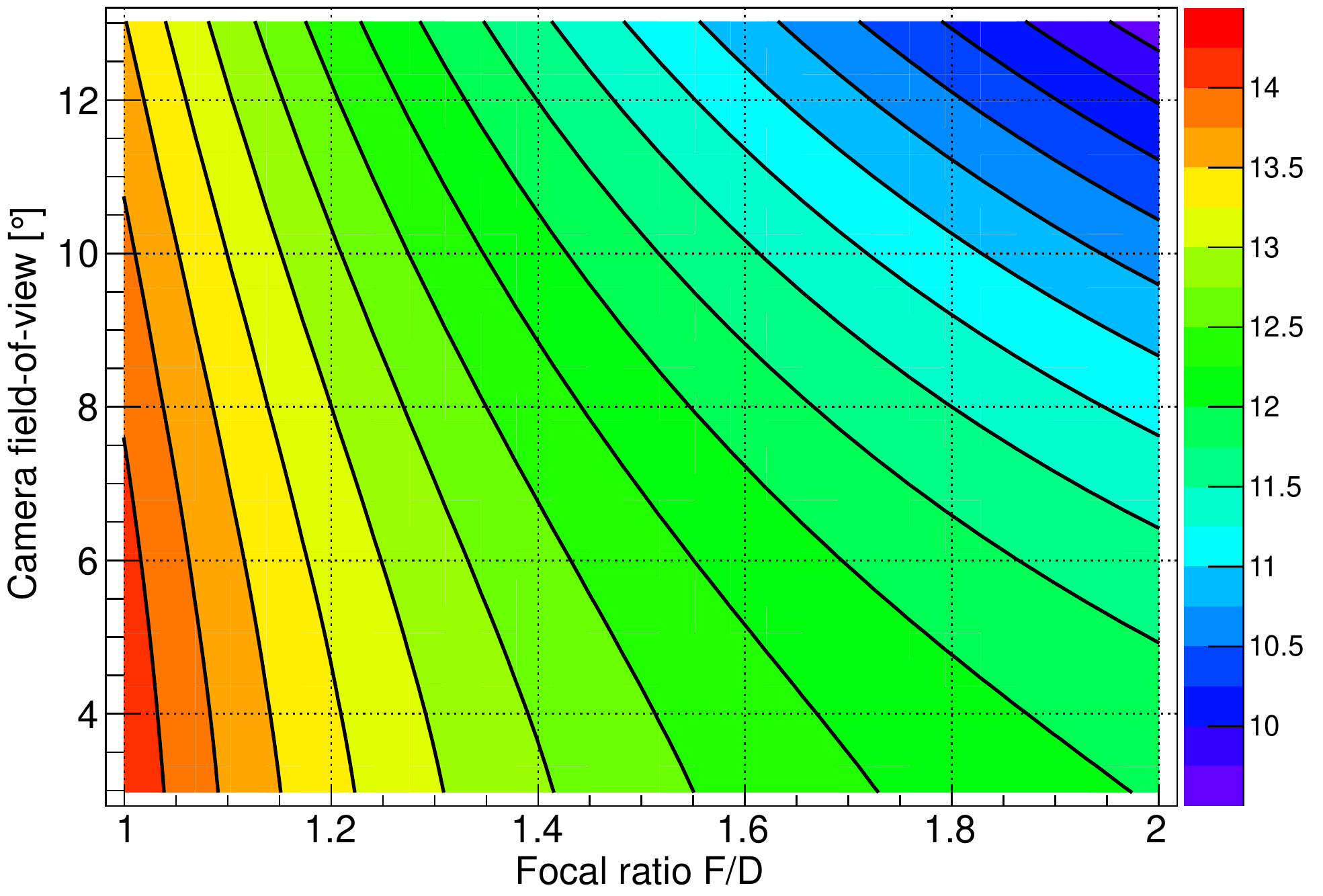}
\caption{The color coded coefficient \(c'\) describing the scale of the
night-sky-background rate versus focal ratio \(F/D\) and camera field-of-view
(diameter) for fixed cone material and photo detector size.\label{fig:nsb}}
\end{figure}

Another interesting aspect for the final performance of a telescope is
the collection of background photons from the diffuse night-sky
background. Here, Eq.~\ref{eqn:EffectiveArea} leads to an interesting
conclusion. Since the rate of the night-sky background photons per
channel scales with the effective reflective area and the solid angle
corresponding to the field-of-view of the pixels, the night-sky
background rate \(R\) is proportional to
\[R\propto A_{eff}\cdot4\pi\sin^2\frac{\vartheta}{2}\approx A_{eff}\cdot\pi\vartheta^2\,,\]
yielding
\[R\propto \frac{\pi^2g}{k}\left(1-\frac{\alpha^2}{g-1}\right)\,. \label{eqn:NSB}\]
For the range of \(F/D=[1,2]\),
Eq.\ref{eqn:FocalLength2} and Eq.~\ref{eqn:f} yield a correction factor
\(g\) between 1.0625 and 1.25. With them,
Eq.~\ref{eqn:NSB} can be transformed into \(R\propto c/k\),
with Eq.~\ref{eqn:SystemConstant} into \(R\propto c'A'n'^2\).
It is immediately apparent that the night-sky background rate scales
with the physical entry area of the pixel.
Assuming only reasonable camera
field-of-views between 3\textdegree{} and 13\textdegree{} diameter,
the coefficient \(c'=2/\sqrt{3}c\) is between 9.6 and 14.2. This can be
interpreted such that the
night-sky background rate per pixel can be considered constant within
\(\pm 10\%\) in the first order along the lines of an optimized
telescope. The dependence of \(c'\) on \(F/D\) and the camera field-of-view
is shown in Fig.~\ref{fig:nsb}.

Quantitatively the night-sky background rate \(r_{\rm{NSB}}\) is given by
$$r_{\rm{NSB}}(r_{\rm{pixel}},\,A_{\rm{mirror}},\,\chi(\nu),\,
f(\nu))=\pi r^2 A_{\rm{mirror}} \int {\rm{d}}\nu\,\frac{f(\nu)}{E_\nu}
\chi(\nu)$$ where \(\chi(\nu)\) and \(f(\nu)\) are the photo-sensor's
photo-detection efficiency and the night-sky background intensity,
respectively. For silicon photo detectors
(MPPC~\cite{hamamatsu_s10985}), as used in the FACT camera, and the
night-sky background at La Palma~\cite{nsb-lapalma}, \(r_{\rm{NSB}}\)
is about 150\,MHz (much larger than the device dark count rate) given a
reflective area of 10\,m\(^2\) and a pixel field-of-view of
0.2\textdegree. If a cutoff in the photo-detection efficiency is
introduced at 650\,nm (PMT-like behavior), this can be further reduced.
In general, the night-sky background rate is not a main problem in
Cherenkov astronomy, as the combined trigger requirements of signal
among nearest pixel neighbors and within a short time lead to its very
efficient suppression.

\paragraph{General considerations}


Existing telescopes are usually under-designed, \ie the photo-detectors
are larger than necessary or the light-concentrators do not reach the
maximum possible concentration.

On the contrary, currently so-called Silicon Photo-multipliers have
proven their potential in Cherenkov telescopes \cite{FactGamma}. These
silicon based photo-detectors usually have a very limited area, but
used in an optimized setup, their effective physical light collection
area, \ie entry of the light concentrator, can be much larger.
If prices of photo-sensors are compared, this has to be taken into
account. Not the price per mm\(^2\) physical sensitive area, but the
price per cone entrance area or field-of-view, has to be considered. If
cheap enough, the signal of several photo-detectors, equipped with
individual light concentrators, could even be summed.

For a comparison, in terms of effective reflective area, the
transmission losses of solid cones and their gain from avoiding Fresnel
reflection has to be taken into account, as well as the reflection
losses of hollow cones.

\paragraph{A note on timing}
For an ideal Davies-Cotton reflector, the arrival time distribution of an instantaneous parallel beam flash is practically flat. More precisely, it is linearly decreasing, but looks flat on the small interval. The number of photons in the arrival time interval \([T, T+\delta T]\) is \(N(T, \delta T)\propto T\). Its width \(\delta T\) is given by \(D/c \cdot f_{DC}(0.5, 0)\), where \(c\) is the speed of light. The interval is of the order 1.1\,ns for a 4\,m class reflector (\(F/D\sim1.5\), up to slightly less than 4.5\,ns for a 12\,m reflector considering \(F/D\sim1.2\). This short time spread is not a problem for the observation of showers with a small size telescope as it is still small compared to the Cherenkov light flash duration. For medium and large size telescopes, a slightly different  mirror arrangement should be chosen if time spread matters. By a mirror arrangement, intermediate between a spherical (Davies-Cotton) and a parabolic design, the time spread can considerably be improved, maintaining the point-spread function almost completely. While the point-spread function is dominated by the majority of the mirrors, i.e. outermost mirrors, the time spread is dominated by the ones with the largest \(Delta T\)  mirrors, i.e. innermost mirrors. Consequently, moving the innermost mirrors closer to a parabola immediately improves the time-spread while the effect on the point-spread function is rather limited. Ideally,  mirrors on a parabola with adapted focal lengths are used, but might be a cost issue. With adapted focal lengths, all mirrors are placed at correct focal distance, so that, a similar point-spread function than for the Davies-Cotton arrangement can be expected.

\paragraph{Remarks about CTA}

\begin{figure*}[htb]
\centering
 \includegraphics*[width=0.48\textwidth,angle=0,clip]{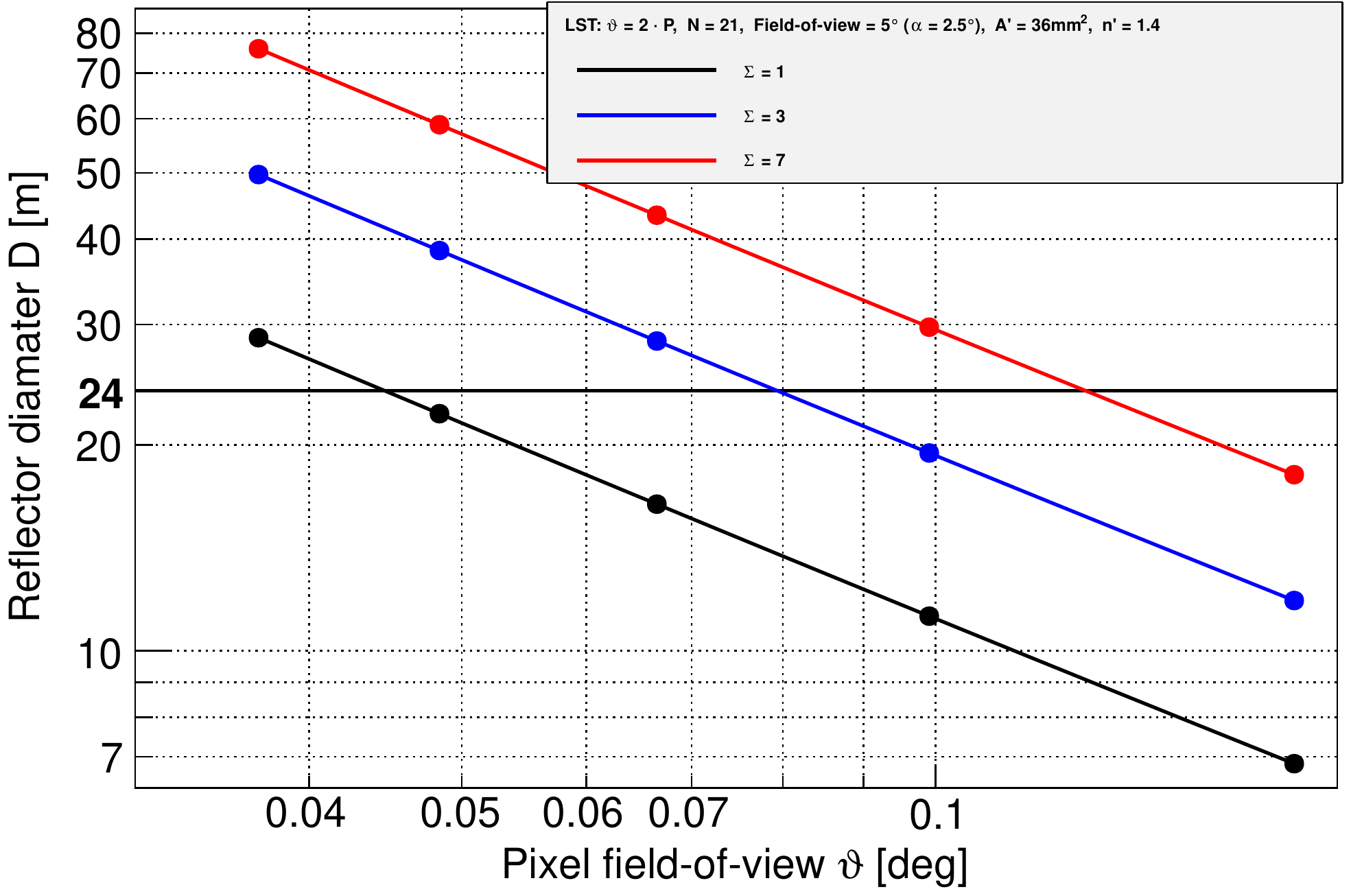}
 \includegraphics*[width=0.48\textwidth,angle=0,clip]{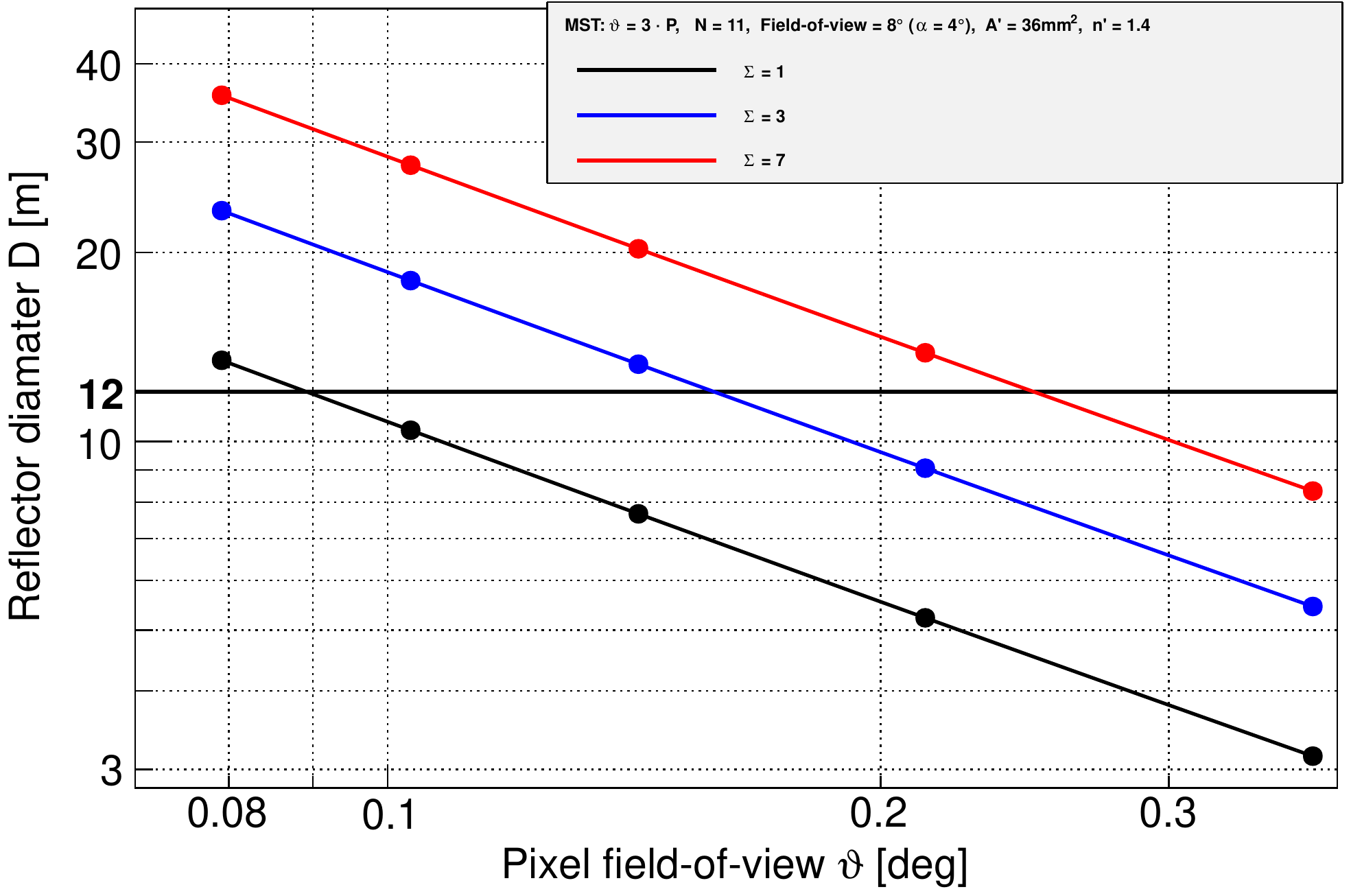}\\\vspace{1em}
 \includegraphics*[width=0.48\textwidth,angle=0,clip]{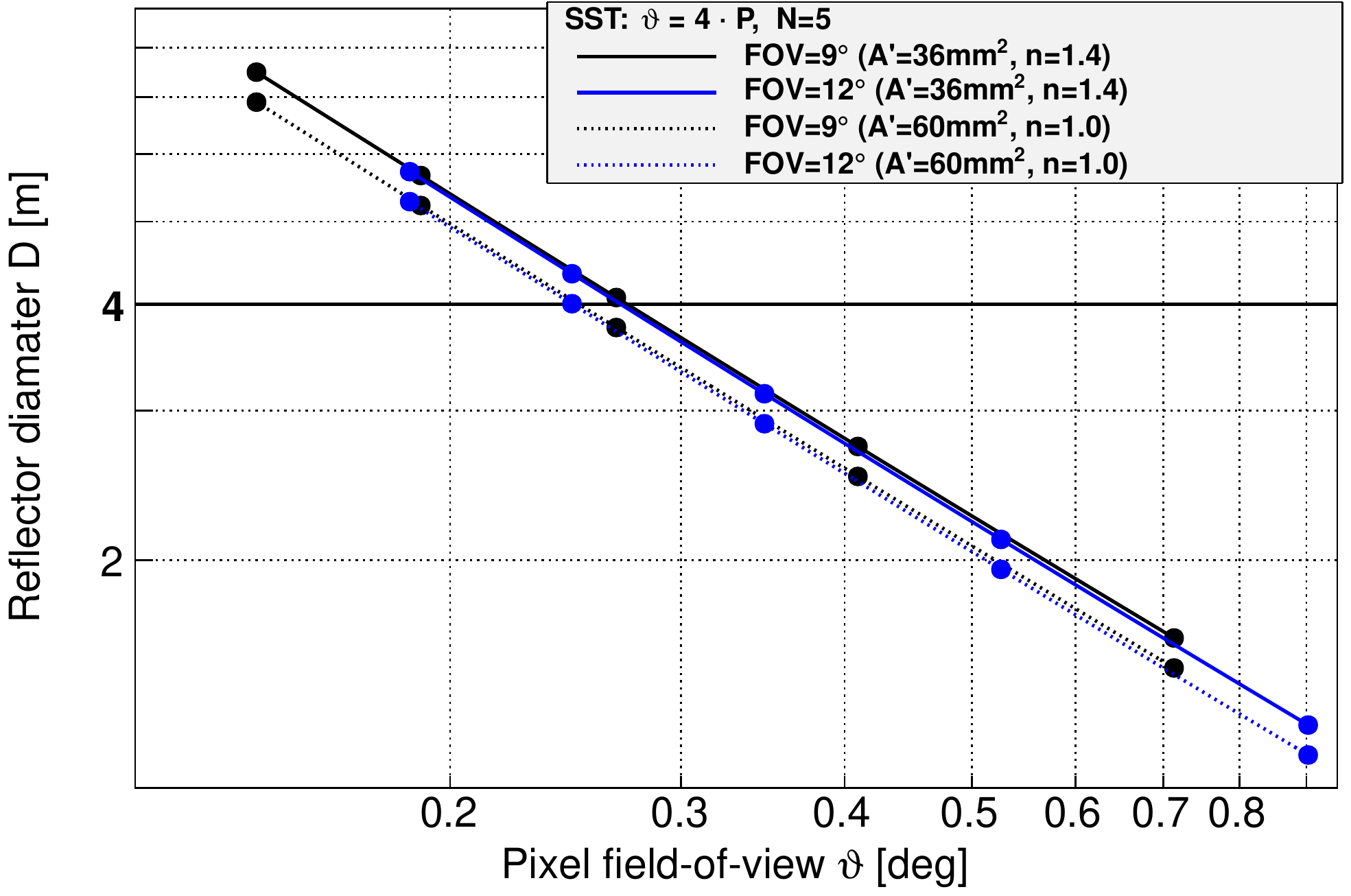}
\caption{Reflector diameter versus pixel's field-of-view for selected 
setups. The shown setups were chosen such that a small-size,
medium-size and large-size telescope is considered. The horizontal
black lines are to guide the eye to possible setups like D=3.5\,m or
D=5\,m, D=12\,m and D=24\,m. The displayed range is between
\(F/D\)\,=\,[1,2]. The dots denote intermediate results at \(F/D\)\,=\,\{1.25,
1.5, 1.75\}. The Greek letter
\(\Sigma\) denotes the number of photo sensors compiled into a single
pixel.
\label{fig:CTA}} 
\end{figure*}

Recent results of FACT~\cite{FactGamma} show that a reflector in
the order of 3.5\,m diameter can give already reasonable physics results
with current analysis and detector technology. Therefore, a 4\,m
diameter reflector for SST is assumed. For physics reason, the
field-of-view is supposed to be between 9\textdegree{} and
12\textdegree{} (leading to a reasonable \(F/D\) between 1.5 and 1.8
assuming an optical quality of 4 and a tessellation number of 5). Requiring a
pixel field-of-view in the order of 0.26\textdegree, possible solutions
could be solid cones with a 36\,mm\(^2\) G-APD or hollow cones with a
60\,mm\(^2\) G-APD, see also Fig.~\ref{fig:CTA}.

The manufacturing of 60\,mm\(^2\) G-APDs is under discussion with
Hamamatsu. A rough estimate shows that a solid cone for such a device
would be about three times longer than for a 9\,mm\(^2\) as used in
FACT. Considering the transmission loss of 10\% in the FACT
cones~\cite{proceeding-IBraun}, which is a very conservative estimate,
such cones would have a loss in the order of 35\%. Since solid cones
avoid the loss from Fresnel reflection at the sealing surface and the
G-APD surface, the real light loss would only be around 27\% assuming
that the hollow cone has a reflectivity of 100\% which in reality is
not true. 

Keeping the pixel field-of-view constant, the gain in reflective area
corresponds to the refractive index of the cone material squared. In
the case of a refractive index of typical Poly(methyl methacrylate) 
PMMA of 1.4 this is a gain of
\(\approx\)100\% reflective area, which outperforms the transmission
loss significantly.

Assuming that the manufacturing of a 36\,mm\(^2\) G-APD would be as
easy as of a 60\,mm\(^2\) G-APD, one can compare a solution with a
36\,mm\(^2\) G-APD and a solid cone and a 60\,mm\(^2\) hollow cone
(assuming perfect reflectivity). In this case, the transmission
loss of the solid cone is around 13\% compared to 8\% Fresnel loss for
the hollow cone. On the other hand, the solution with the hollow cone
yields a 15\% smaller reflective area (same pixel field-of-view) or
27\% more pixels (same reflective area). 

Assuming further that the price of the camera scales with the 
price of each channels, a reduction of the number of channels by almost
30\% reduces the costs for the camera significantly. Since the costs
are also dominated by the price for the photo-detectors, and the price of
G-APDs, in the first order, scales with the sensitive area, it can be
estimated that the price for the 36\,mm\(^2\) G-APDs would be almost 
a factor of two lower than for the larger ones. 

In Figure~\ref{fig:CTA} possibly solutions for MST and LST designs are shown
using G-APDs and solid cones. On both cases it is convenient to sum at
least three, or even seven, pixels into one readout channel to keep
the ratio \(F/D\) low for construction reasons.

\section{Conclusion}
\begin{figure}[htb]
\centering
 \includegraphics*[width=0.48\textwidth,angle=0,clip]{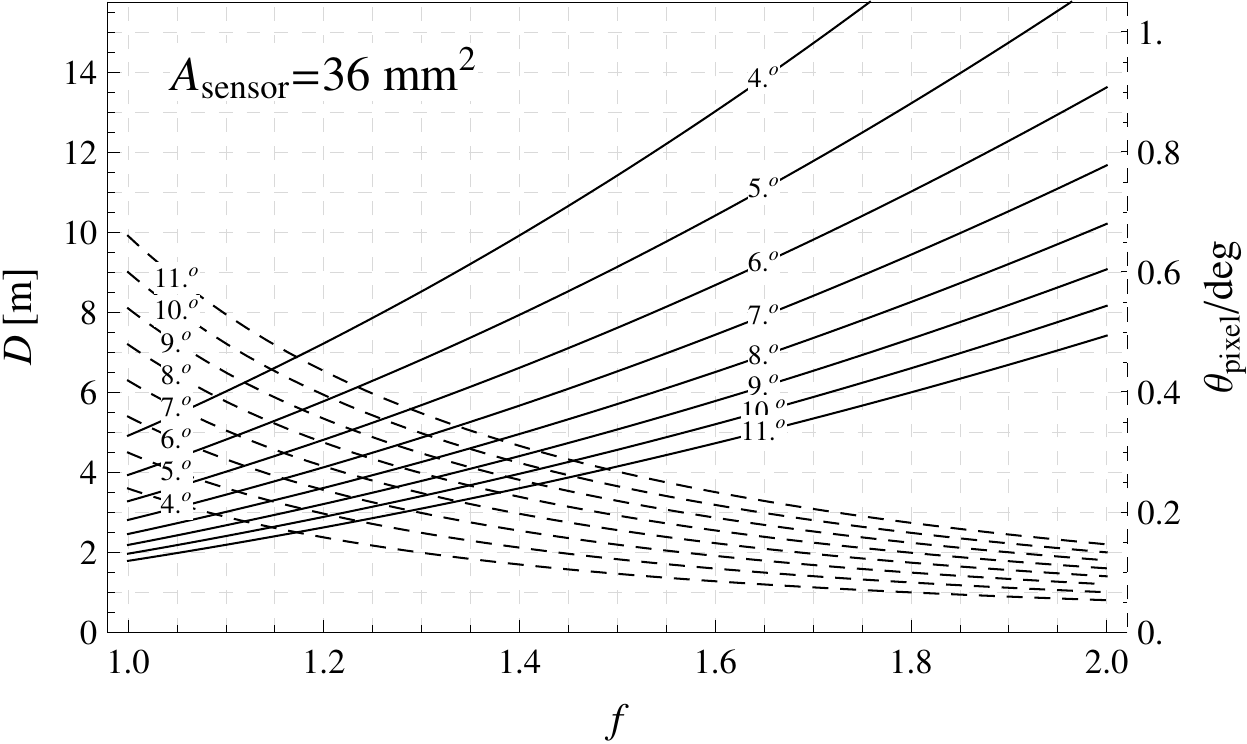}
\caption{Example design overview for a small size telescope calculated
for an ideal Davies-Cotton reflector, \ie \(N\sim9\), a pixel size
four times the point-spread function and solid cones with \(n=1.4\).
For a given \(F/D\) and camera
field-of-view, the corresponding reflector diameter and the pixel field-of-view
can be read.
\label{fig:Conclusion}} 
\end{figure}

The Davies-Cotton design with its simplicity as compared to non
validated dual optic systems is assuredly a good option for a wide
field-of-view, up to 10\textdegree\,--\,12\textdegree, high energy
Cherenkov telescope.

With this study, it is possible to scan a wide
phase space of the design of Cherenkov telescopes or telescope arrays. This was
achieved by a description of the optical performance of
Davies-Cotton reflectors and introduction of the effect of light-concentrators.
In particular, this study provides an analytical description of the
optical performance of a tessellated Davies-Cotton reflector precise enough
to enable performance studies without the need for dedicated simulations.

By including the effect of the light-collector into the system of
equations, the available phase space of design parameters
is reduced to a single parameter, once the photon detector has been chosen
and either the pixel field-of-view or the camera field-of-view has
been fixed by physics constraints. 
While the choice of photo sensor is usually defined by the availability
on the market, constraints on the camera field-of-view are a result of the
physics targets.

If these two parameters are fixed, the whole available phase space of
possible solution can now be scanned by changing a single input parameter.
It can, for example, be convenient to scan a reasonable range of the focal
ratio \(F/D\) and derive all other parameters accordingly.
From the result, the most cost efficient solution, or the one performing best in
sense of physics targets can be chosen.

For the Cherenkov telescope array, several design options were presented.
It could be shown that for the small size telescope, considering a camera
field-of-view of 9\textdegree{} to 12\textdegree, a four meter reflector
is enough if 36\,mm\(^2\) senors are topped with solid cones
to achieve a pixel field-of-view in the order or 0.25\textdegree{} to 0.3\textdegree{}
at reasonable \(F/D\). An alternative solution are hollow cones with correspondingly
larger sensor area, which is disfavored because of the costs dominated by the
sensor. An example plot which easily allows to determine reasonable options
from the available phase space is shown in Fig.~\ref{fig:Conclusion}. The reflector 
diameter can easily be re-scaled linearly with the photo sensor size and 
the refractive index of the cone material.

For the medium size and large size telescopes, the most reasonable solution
using small sensors would be the summation of three and seven, respectively. 
Equipped with different sum-stages, these modules could be applied in any
telescopes. Larger silicon based sensors, expected soon on the marked, would
allow a single-channel/single-sensor solution. Using several small sensors 
in one channel has the advantage that the application of solid cones is possible
in terms of weight and transmission and costs for photo sensors can be kept 
low due to their at least two times higher compression ratio.




\newpage
\appendix

\section{Parametrizing a Davies-Cotton reflector}

\subsection{Ideal Davies-Cotton reflector}\label{appendix:A}

The Davies-Cotton design \cite{dc-design} is known to be promising for
wide field prime-focus telescopes and was studied earlier
analytically~\cite{Vassiliev} and through simulations~\cite{mirzoyan}.
However, the parametrizations are moderately accurate and non existing
for tessellated reflectors.

Here, parametrizations are provided, accurate at the percent level up
to 12\textdegree{} field-of-view, for the ideal (non-constructable)
Davies-Cotton telescope and accurate to a few percent for a realistic
Davies-Cotton telescope with arbitrary tessellation of the reflector.

\paragraph{Prime-focus telescope design}
The major issue of the design of a telescope is the reflector and its 
optical performance. Since design parameters like the field-of-view
of a single pixel or the field-of-view of the whole camera are closely
related to the reflectors optical performance, it is important 
to understand the relation between the reflector design and its
performance. 
Unfortunately, neither spherical nor parabolic mirrors can provide both,
good optical point spread function for on-axis and inclined rays, 
at the same time, because
the distance between any point on the mirror surface to the focal point
does not match the local focal length defined by the local
radius of curvature.
Furthermore, in the case of a spherical mirror, also the shape of
the mirror surface is not ideal compared to a parabolic mirror. 
The parabolic shape ensures that parallel rays from infinity are
well focused into a single point (due to the definition of a 
parabolic surface) while in the spherical case this is not the case.
That means that in both cases rays hitting the mirror far off its
center have their focal point not at the focal plane. In the case of a
spherical mirror they also miss the focal point ({\it aberration}).

Consequently, the ideal mirror would be a combination of two
properties: A mirror surface which is shaped such that it has the
right focal distance at any point, but at the same time any point
is correctly oriented, so that focal distance and direction are correct.
Since local normal vector and local curvature cannot be disentangled
such a mirror can only be a theoretical construction.
Tessellating the reflector into individual mirrors, this behavior can be
approximated, as shown by Davies and Cotton, if the reflector is built
from several spherical mirrors which are placed on a sphere around the
focal point. In this case, the reflector can have the correct focal
distance locally and, at the same time, the mirror elements can be
oriented such that they correctly focus to the focal point.
Apart from an improved optical performance for inclined rays, 
the production of several small and identical mirrors is also
much more cost efficient than the production of a single large mirror.

Since any optical system can always be linearly scaled, in the
following a scale factor is chosen such that the reflector diameter
corresponds to unity, which is identical to defining \(f=F/D\) with
\(F\) being the focal length and \(D\) the diameter of the mirror.

\paragraph{Spherical reflector} 
The spherical mirror has its focal point at half its radius of
curvature \(f\). Its surface is given by \(z=f_{\rm{sph}}(x,y)\) and its
normal vector by \(\vec {n}_{\rm{sph}}(x,y)\):
\begin{eqnarray} 
f_{\rm{sph}}(x,y)&=& 2f - \sqrt{(2f)^2 - (x^2 + y^2)} \\
\vec {n}_{\rm{sph}}(x,y) &=& (x/(2f - f_{\rm{sph}}),\ y/(2f - f_{\rm{sph}}),\ -1)
\end{eqnarray} 

\paragraph{Ideal Davies-Cotton reflector} 
The ideal Davies-Cotton reflector has a non constructable surface. Its
shape \(z=f_{\rm{DC}}(x,y)\) is spherical with radius of curvature \(f\),
but its normal vectors are defined to intercept at location
\(\vec{F}=(0,0,2f)\). Formally, the surface equation \(f_{\rm{DC}}(x,y)\) and
the surface normal vectors \(\vec {n}_{\rm{DC}}(x,y)\) are  
\begin{eqnarray} 
f_{\rm DC}(x, y) &=&  f - \sqrt {f^2 - (x^2 + y^2)} \\ 
\vec {n}_{\rm DC}(x,y) &=& \nabla (f_{\rm DC}(x, y) - z) \\ \nonumber
&=& \left(\frac{x}{2 f - f_{\rm DC}(x, y)},\ \frac{y}{2 f - f_{\rm DC}(x, y)},\ -1 \right)
\end{eqnarray}

Practically, this equations describes infinitely small mirror elements
placed on a sphere, oriented accordingly.


\paragraph{Taylor development} 

To have the root-mean-square of the projection of reflected rays on the focal plane
along \(x\) and \(y\) coinciding with tangential and sagittal
resolutions, a rotation around \(z\) is performed without
loss of generality.

An incoming ray with vector \(\vec{v}=(0,\ \sin\phi,\cos\phi)\) will
therefore be reflected on the surface in the direction  
\(\vec{v}_r=\vec{v}-2(\vec{v}\cdot\vec{n})\vec{n}/n^2\) and intercept the (non curved)
focal plane at \(r=\sqrt{x^2+y^2}\), generally yielding
%
%
%
\[ (X, Y, Z) = \left(x + \frac{v_ {rx}}{v_ {rz}}(f-f(x,y)),\ y + \frac{v_ {ry}}{v_ {rz}}(f-f(x,y)),\ f \right)\,. \]
For the ideal Davies-Cotton this takes the explicit form
\[ (X, Y, Z)_{\rm DC} = \left(x + \frac{v_ {rx}}{v_ {rz}}\sqrt {1 - \frac{r^2}{f^2}},\ y + \frac{v_ {ry}}{v_ {rz}}\sqrt {1 - \frac{r^2}{f^2}},\ f \right)\,. \]

It is straightforward to numerically calculate the image centroid
\((\bar\xi,\bar\eta)\) and the resolution \((\Delta\xi, \Delta\eta)\) of
such a telescope and to estimate the contribution of various terms to
the resolution with a Taylor development of \(X\) and \(Y\) in terms of
\(x\), \(y\) and \(\phi\).
The development of terms of the form \(x^i y^j \phi^k\), with
\(i+j\le5\) and \(k\le3\) was found to be sufficient for a percent
precision in the resolution parameters.

The tangential and sagittal barycenter in the focal plane (the image centroid) for a uniform beam on the primary surface are given by
\begin{eqnarray}
\bar\xi &=& \frac{\int_ 0^{1/2} r\,\rm{d}r \int \rm{d}\theta \, Y (x = r\cos\theta, y = r\sin\theta)}{\int_ 0^{1/2} r {\rm{d}}r \int {\rm{d}}\theta}\,,\nonumber\\
\bar\eta &=& \frac{\int_ 0^{1/2} r\,\rm{d}r \int \rm{d}\theta \, X (x = r\cos\theta, y = r\sin\theta)}{\int_ 0^{1/2} r {\rm{d}}r \int {\rm{d}}\theta}\,,\nonumber\\
&&
\end{eqnarray}
and the corresponding resolution in term of root-mean-square:
\begin{eqnarray}
\Delta\xi^2 &=& \frac{{\int_ 0^{1/2} r\,\rm{d}r \int {\rm{d}}\theta \, (Y (x = r\cos\theta, y = r\sin\theta) - \bar\xi)^2}}{{\int_ 0^{1/2} r {\rm{d}}r \int {\rm{d}}\theta}}\,,\nonumber\\
\Delta\eta^2 &=& \frac{{\int_ 0^{1/2} r\,\rm{d}r \int {\rm{d}}\theta \, (X (x = r\cos\theta, y = r\sin\theta) - \bar\eta)^2}}{{\int_ 0^{1/2} r {\rm{d}}r \int {\rm{d}}\theta}}\,.\nonumber\\
&&\label{expr9}
\end{eqnarray}
The upper integration bound in \(r\) originates from the fact that the
optical system was scaled to meet a reflector mirror of \(d\)\,=\,1, hence
\(r\)\,=\,\(1/2\).

Taylor development of the above formulas brings the desired result
\begin{eqnarray}
\Delta\xi^2&=&\frac{1}{2^4}\sum_{i, j} \frac{c^\xi_{i, j} }{f^j} \phi^i\,,\label{eq:DeltaXi}\\
\Delta\eta^2&=&\frac{1}{2^4}\sum_{i, j} \frac{c^\eta_{i, j}}{f^j} \phi^i\,.\label{eq:DeltaEta}
\end{eqnarray}
Coefficients given in Table~\ref{tab:coeff}. For the ideal
\begin{table}
\begin{center}
\begin{tabular}{|c|cc|cc|}\cline{2-5}
\multicolumn{1}{c|}{}& sph. & DC  & sph. \nth{3} & DC approx.            \\
\multicolumn{1}{c|}{}&  &  &  order~\cite{mirzoyan} & from~\cite{Vassiliev} \\  \hline
\(c^{\xi, \eta}_{0,6}\) & $\null{\frac{1}{2^{11}}}$         & 0 & $\null{\frac{1}{2^{11}}}$ & 0   \\
\(c^{\xi, \eta}_{0,8}\) & $\null{\frac{3^2}{2^{14} 5}}$     & 0 & 0 & 0\\
\(c^{\xi, \eta}_{0,10}\)& $\null{\frac{7\cdot 13}{2^{23}}}$ & 0 & 0 & 0\\
\hline\hline
\(c^\xi_{2,4}\) & $\null{\frac{7}{2^5 3}}$                   &  $\null{\frac{1}{2^{6}}}$     & $\null{\frac{7}{2^5 3}}$ & $\null{\frac{1}{2^{6}}}$  \\
\(c^\xi_{2,6}\) & $\null{\frac{131}{2^{11} 3}}$              & $\null{\frac{5}{2^{10}}}$     & 0 &   $\null{-\frac{1}{2^{8}}}$        \\
\(c^\xi_{2,8}\) & $\null{\frac{257\cdot 307}{2^{19} 3^2 5}}$ & $\null{\frac{191}{2^{16} 5}}$ & 0 & 0 \\
\hline\hline
\(c^\xi_{4,2}\) & \multicolumn{2}{c|}{1} & 1 & 1 \\
\(c^\xi_{4,4}\) & $\null{\frac{7}{2^4}}$           & $\null{\frac{5\cdot 7}{2^5 3}}$     & 0 & $\null{\frac{5\cdot 7}{2^5 3}}$       \\
\(c^\xi_{4,6}\) & $\null{\frac{1163}{2^{10} 3^2}}$ & $\null{\frac{5\cdot 67}{2^{10} 3}}$ & 0 &  0  \\
\(c^\xi_{4,8}\) & $\null{\frac{19\cdot 67\cdot 1039}{2^{19} 3^3 5}}$ &  $\null{\frac{8527}{2^{15} 3\cdot 5}}$   & 0 &  0 \\
\hline\hline
\(c^\eta_{2,4}\)  & \multicolumn{2}{c|}{$\null{\frac{1}{2^5 3}}$} & $\null{\frac{1}{2^5 3}}$ & $\null{\frac{1}{2^5 3}}$        \\
\(c^\eta_{2,6}\)  & $\null{\frac{1}{2^{8}}}$       &  $\null{\frac{3}{2^{10}}}$   & 0 & 0 \\
\(c^\eta_{2,8}\)  & $\null{\frac{19}{2^{12} 5}}$   & $\null{\frac{29}{2^{13} 5}}$ & 0 & 0  \\
\(c^\eta_{2,10}\) & $\frac{5\cdot 1459}{2^{24} 3}$ & $\frac{5\cdot 53}{2^{21}}$   & 0 & $\frac{3}{2^{10}}$ \\
\hline\hline
\(c^\eta_{4,4}\) & \multicolumn{2}{c|}{ $\null{\frac{1}{2^4 3^2}}$ } & 0 & $\null{\frac{1}{2^4 3^2}}$ \\
\(c^\eta_{4,6}\) & \multicolumn{2}{c|}{ $\null{\frac{5}{2^{10}}}$ }  & 0 & 0 \\
\hline
\end{tabular}
\end{center}
\caption{\label{tab:coeff}Table of coefficients $c_{i,j}$ for Eq.~\ref{eq:DeltaXi} and \ref{eq:DeltaEta}}
\end{table}
Davies-Cotton, only leading terms \(\frac{c^\xi_{i, j}}{f^j} \phi^i\) and
\(\frac{c^\eta_{i, j}}{f^j} \phi^i\) are retained, at \(f=1\) and a maximum off-axis
angle of the incoming rays of \(\phi_{\rm{max}}\)=5\textdegree, \(ij\)
terms such that \(c^\xi_{i, j}\phi^i/(2f^j\Delta\xi)>10^{-3}\) and
\(c^\eta_{i, j}\phi^i/(2f^j\Delta\eta)>10^{-3}\).

%

As we apply the same conditions to spherical prime-focus design, less
terms are present at higher order, \ie less spherical aberration.
To mirror the result for the ideal Davies-Cotton, several non leading
terms are added giving a consistent picture for both developments as
shown in Table~\ref{tab:coeff}.

\paragraph{Comparisons} 

\begin{figure}[htb]
\centering
 \includegraphics*[width=0.47\textwidth,angle=0,clip]{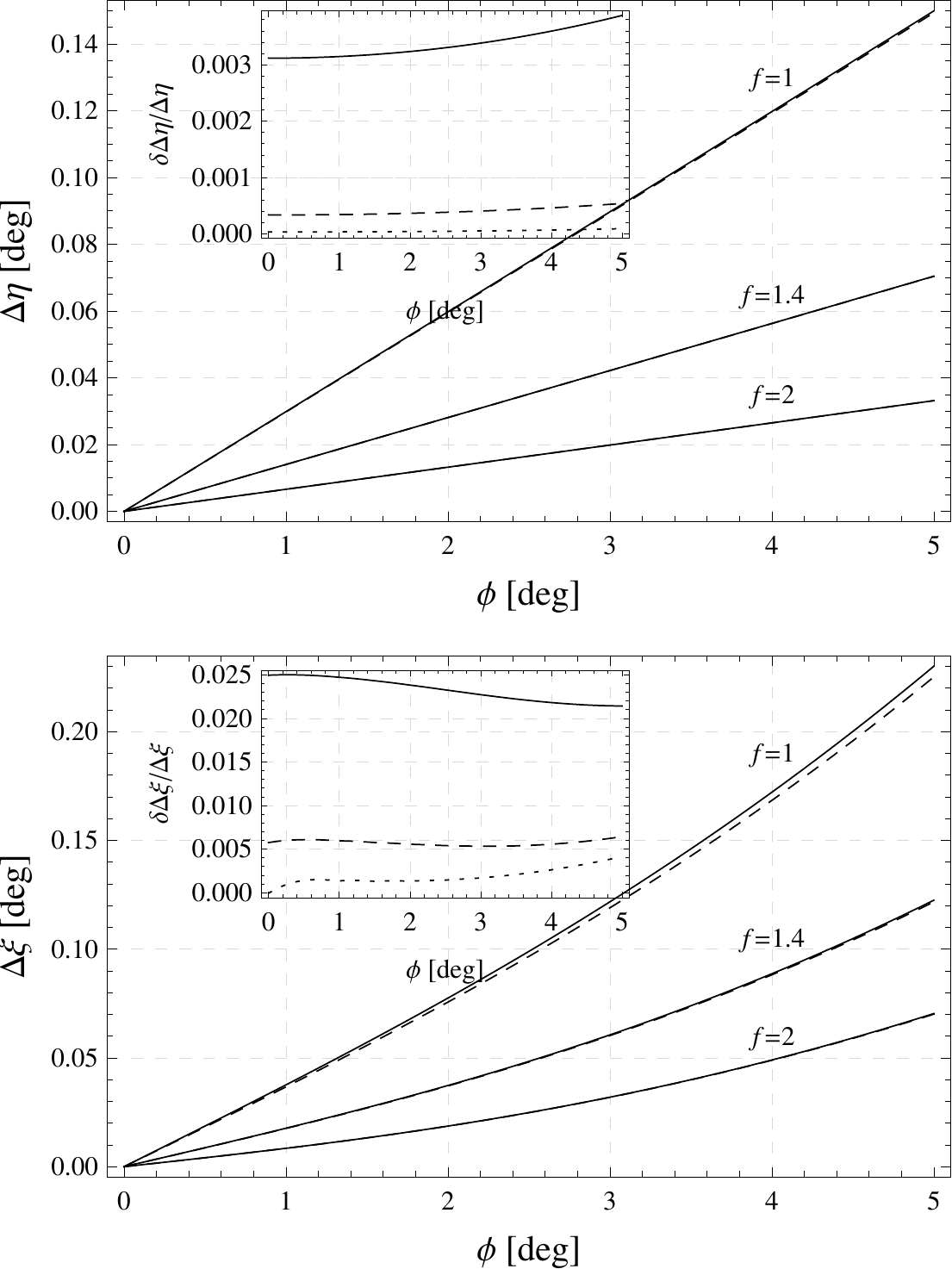}
\caption{Comparison between the exact (solid), \ie numerically calculated, resolution
parameters \(\Delta\xi\) and \(\Delta\eta\) and the result of the limited Taylor
development (dashed). 
The inset shows the ratio of both for \(f=\{1.0, 1.4, 2.0\}\) (solid, dashd, dotted).
}
\label{fig:DCtelReso-approx}
\end{figure}

A comparison between exact (numerically calculated) results and the
presented limited Taylor development for both designs is presented in
Fig.~\ref{fig:DCtelReso-approx}.

In \cite{Vassiliev} and \cite{mirzoyan}, \nth{3} order developments for
the Davies-Cotton and sherical mirror, respectively, have been discussed.
The obtained coefficients are repeated here for
completeness in table~\ref{tab:coeff}. Both solutions show
up to 20\% fractional error \(\delta\Delta\eta\), \eg at \(f=1\).

While in~\cite{mirzoyan}, terms \(x^i y^j \phi^k\) in the development
were kept only to the \nth{3} order, \ie \(i+j\le3\), here terms 
were kept up to \(i+j\le5\) and \(k\le3\).

\begin{figure}[htb]
\centering
 \includegraphics*[width=0.47\textwidth,angle=0,clip]{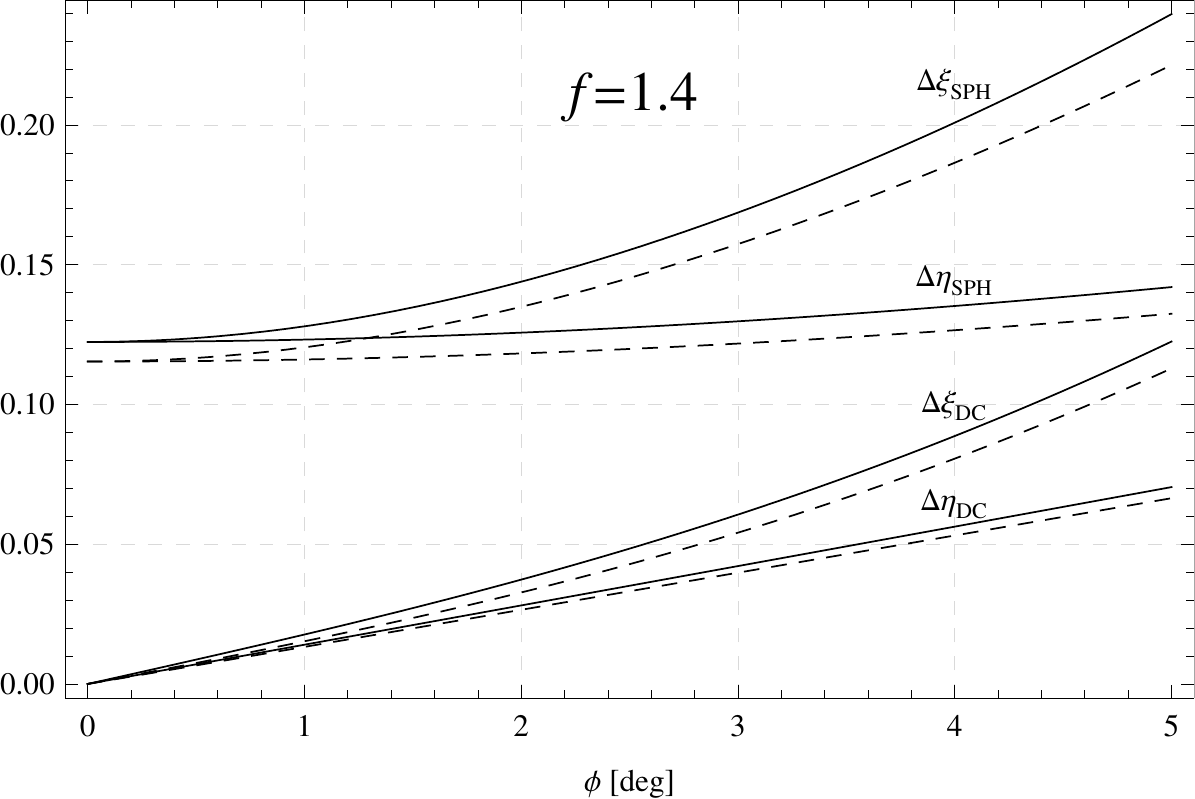}
\caption{Comparison between the exact, \ie numerically calculated,
resolution parameters \(\Delta\xi\) and \(\Delta\eta\) and the result of the limited
Taylor development from~\cite{mirzoyan} and~\cite{Vassiliev}.}
\label{fig:telReso-compare}
\end{figure}

At the expense of the introduction of more terms, consequently, the precision
of the presented development is about ten times better as illustrated in
Fig.~\ref{fig:telReso-compare}. 

\begin{figure}[htb]
\centering
 \includegraphics*[width=0.47\textwidth,angle=0,clip]{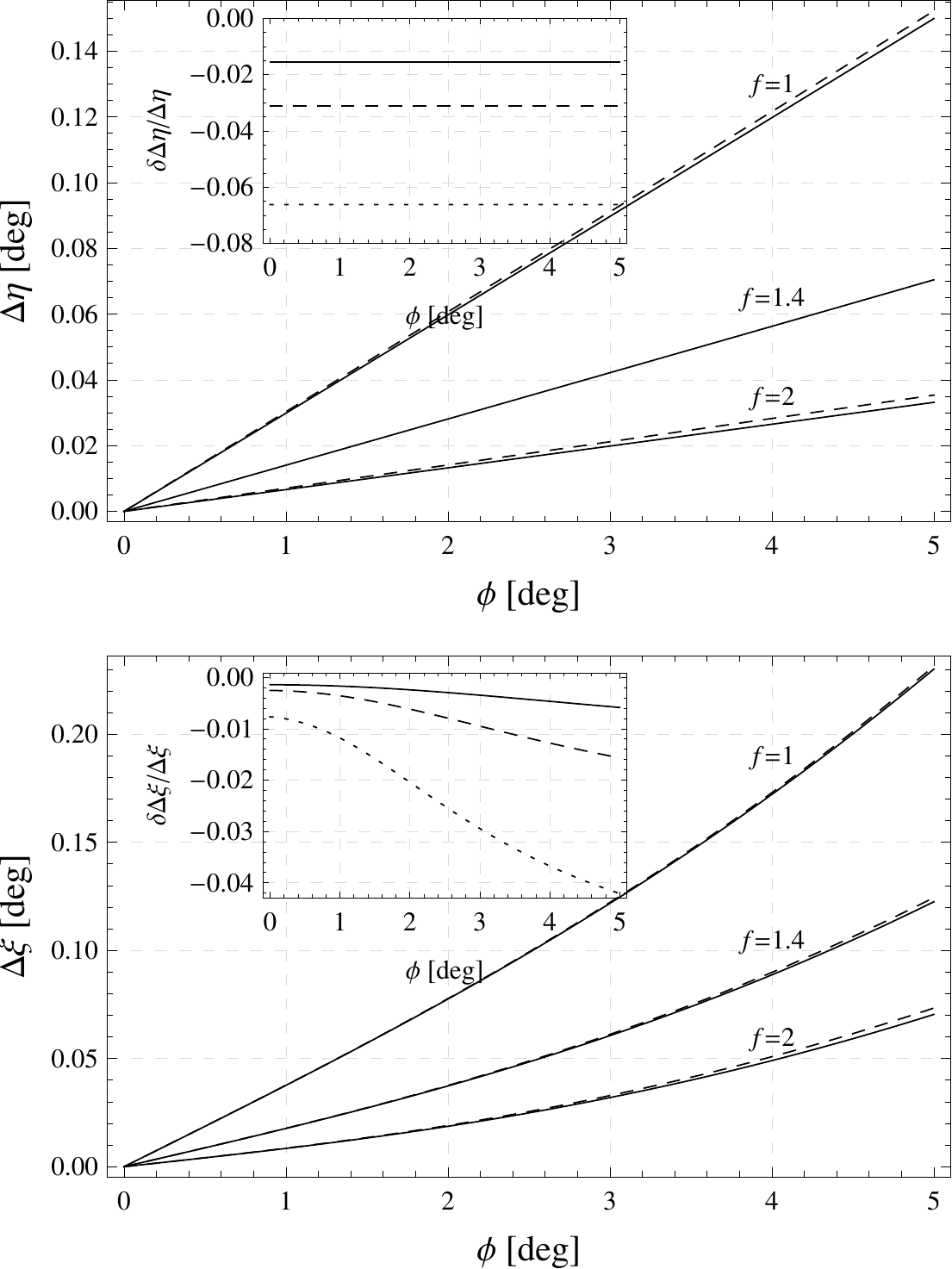}
\caption{Resolution parameters \(\Delta\xi\) and \(\Delta\eta\) of the exact results, \ie numerically calculated, with (dashed) and without (solid) obscuration. Inset shows the ratio of both
for \(f\)\,=\,\{1,1.4,2\} (solid, dashed, dotted).
}
\label{fig:DCtelReso-obscur}
\end{figure}

\paragraph{Obscuration} In the above considerations, the shadow of a
possible detector in the focal plane has been neglected.  By changing
the lower bound from \(r=0\) to \(r=5\)\textdegree\(\cdot f\) in expressions
\ref{expr9}, obscuration can easily be quantified.
Fig.~\ref{fig:DCtelReso-obscur} shows that obscuration degrades the
resolution parameters by about 1.5\% at \(f=1\) up to about 6\% at
\(f=2\), 5\textdegree{} off-axis.

\subsection{Parameterization for a tessellated reflector}\label{appendix:B}
A realistic implementation of the non-constructable Davies-Cotton
telescope consists in introducing a reflector made of multiple
individual spherical mirrors. The tessellation number is defined as the
the number \(N\) of mirrors in the diagonal. In the limit
\(N\)\,=\,\(\infty\), it is identical to the ideal Davies-Cotton design. In
practice, \(N\) is a number \(\lesssim\)30.

\paragraph{Simulation} The effective parametrization is presented as a
correction to the limited Taylor development derived earlier. The correction
is implemented through ray-tracing simulation performed with the MARS
software (described in \cite{mars1, mars2, mars3}), which do fully
reproduce the results obtained earlier in the case of a spherical and
ideal Davies-Cotton reflectors. Although an ideal Davies-Cotton
reflector cannot be build in reality, it can be simulated easily.
Simulations enable the use of arbitrary tessellation, since analytical
solution being not very well suited for this task. 

For the simulation the following properties have been used:
\begin{itemize}

\vspace{-1ex}\item Individual mirrors are hexagonal. For symmetry 
reasons each hexagon is rotated by 15\textdegree{} against the x-/y-axis

\vspace{-1ex}\item The mirrors are fixed on a hexagonal grid in the
\(x\)/\(y\)-plane with spacing \(d\)

\vspace{-1ex}\item The diameter of the individual mirrors is defined as \(nd^2=\pi
D^2/\sqrt{3}\), with \(n\) being the total number of mirrors in the
system

\vspace{-1ex}\item Their center is located on a sphere around the focal point (this
corresponds to \(f_{\rm DC}\) for the ideal Davies-Cotton)

\vspace{-1ex}\item The focal length \(F\) of each mirror is equal to the radius of
the sphere, and therefore equal to the focal length of the system

\vspace{-1ex}\item The overall shape of the reflector is also hexagon like

\vspace{-1ex}\item The tessellation number \(N\) is the number of mirrors in the
diagonal

\vspace{-1ex}\item Each mirror is oriented to a virtual point in 2\(F\) (this
corresponds to \(n_{\rm DC}\) for the ideal Davies-Cotton)

\vspace{-1ex}\item The small effect of obscuration by the focal instrumentation is
neglected

\vspace{-1ex}\item The mirror surface is assumed to be ideal

\end{itemize}

\begin{figure}[htb]
\centering
 \includegraphics*[width=0.3\textwidth,angle=0,clip]{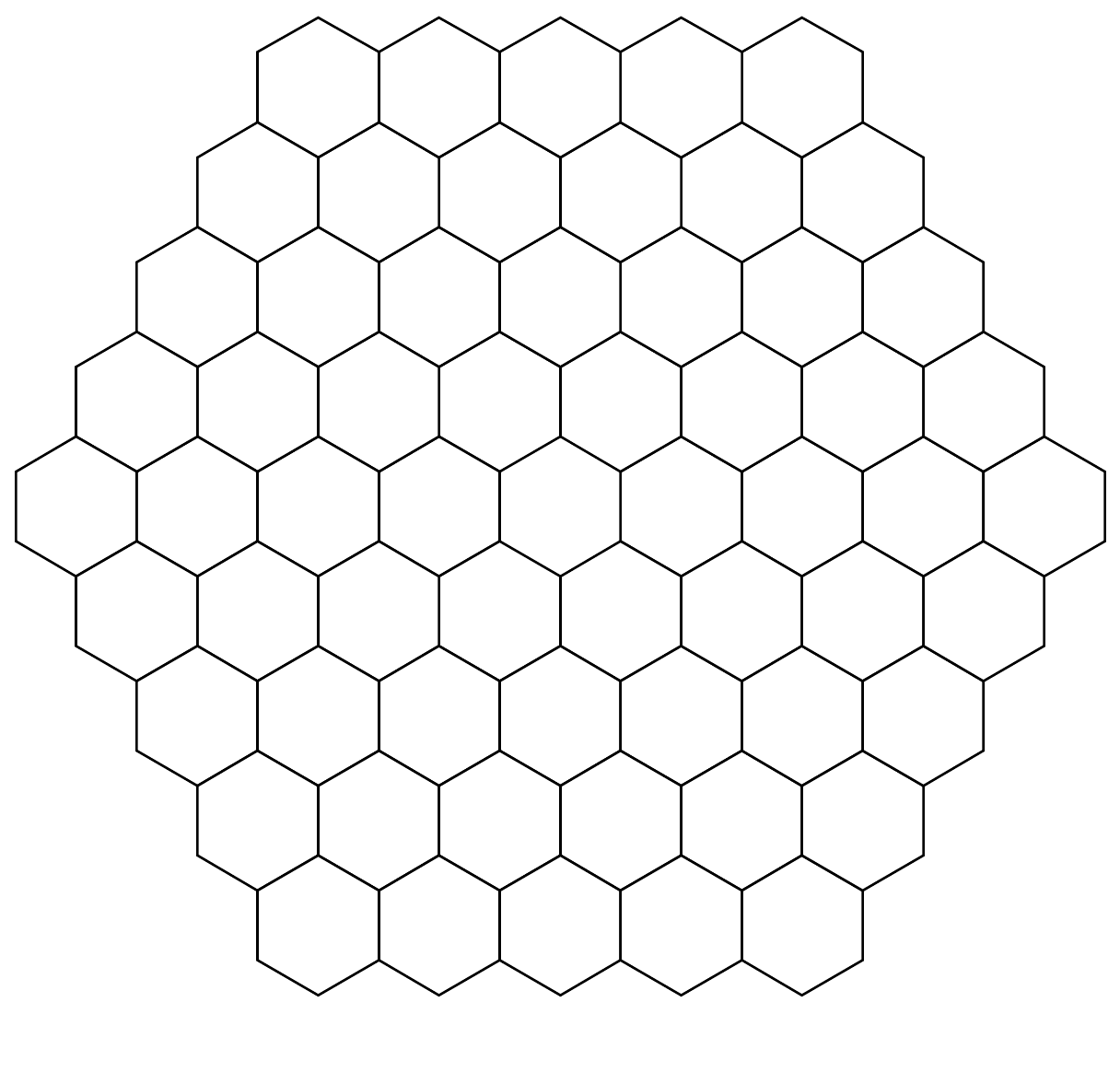}
\caption{An example of a reflector layout for \(N\)=9
mirrors on the diagonal. This also includes \(N\)=\{3, 5, 7\} removing
the rows of outer mirrors consecutively. }
\label{fig:Reflector9}
\end{figure}

An example for such a reflector is given in Fig.~\ref{fig:Reflector9}


Simulations have been carried out for \(N\) between 1 and 79 in steps
of two, in the range \(1 \le F/D \le 2\) in steps of 0.1 and for rays
with off-axis angles comprised in
0\textdegree\(\le\phi\le\)6.5\textdegree{} in steps of 0.5\textdegree.

Empirically, it could be found that introducing a dependence on the
tessellation number, the formulas given in
\ref{appendix:A} for the spherical mirror and the ideal Davies-Cotton
mirror could be unified. For this, a
linear dependence at \nth{0} order in \(\phi\), and a quadratically at order
\(\phi^i\) for \(i\ne0\), has been introduced. Additionally, an
effective rescaling \(f_{\rm eff}=f/w\) is needed to reach an accuracy about 5\%
in the whole simulated range. The root-mean-square of a tessellated
Davies-Cotton can then be written as
%
%
%
\begin{eqnarray}
\Delta\xi^2 &=& \frac{1}{2^4}\left(\sum_{j} \frac{s^\xi_{0,j}}{N^2 f^j_{\rm eff}}
+\sum_{i>0,j} \left[\frac{s^\xi_{i,j}-d^\xi_{i,j}}{N}+d^\xi_{i,j}\right] \frac{\phi^i}{f^j_{\rm eff}}\right)\,, \nonumber\\
\Delta\eta^2 &=&  \frac{1}{2^4}\left(\sum_{j} \frac{s^\eta_{0,j}}{N^2 f^j_{\rm eff}}
+\sum_{i>0,j} \left[\frac{s^\eta_{i,j}-d^\eta_{i,j}}{N}+d^\eta_{i,j}\right] \frac{\phi^i}{f^j_{\rm eff}}\right)\,.\nonumber\\
&&
\label{eq:realDC}
\end{eqnarray}
The coefficients \(s_{i,j}\) and
\(d_{i,j}\) are the ones given in Table~\ref{tab:coeff} for the spherical
and the ideal Davies-Cotton, respectively. 
\begin{figure}[htb]
\centering
 \includegraphics*[width=0.48\textwidth,angle=0,clip]{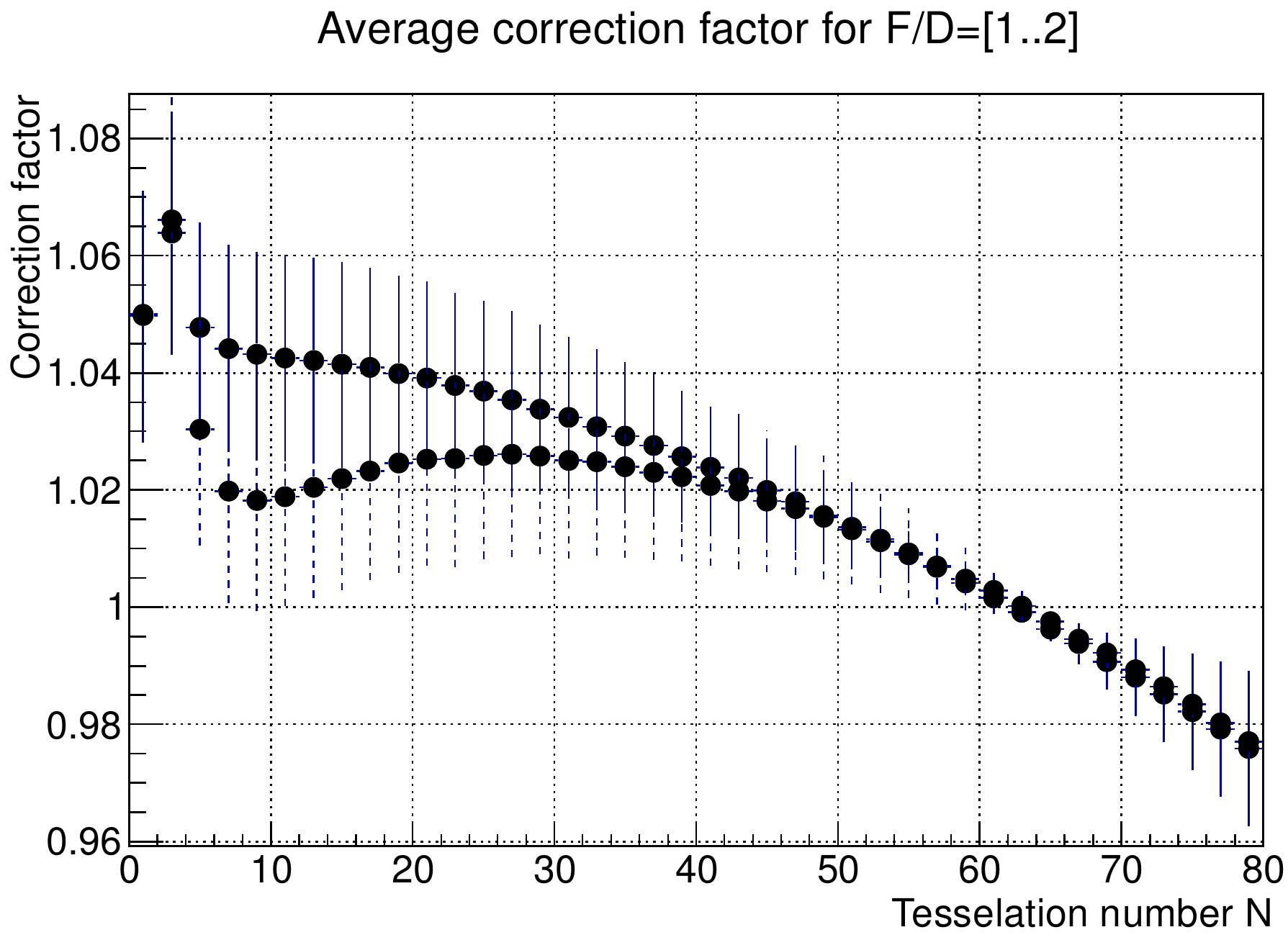}
\caption{The scale factor \(w\) as defined by formula~\ref{eq:realDC}
determined from fits to the simulated point-spread function. Each point
is the average of all correction factors obtained for the simulated
range of \(F/D\). The error bar denotes the spread. The two curves are
the tangential (upper curve, solid) and sagittal (lower curve, dashed),
respectively.}                                           
\label{fig:correction}
\end{figure}

The tessellation number \(N\) can here be interpreted as a parameter
\begin{figure*}[htb]
\centering
 \includegraphics*[width=0.48\textwidth,angle=0,clip]{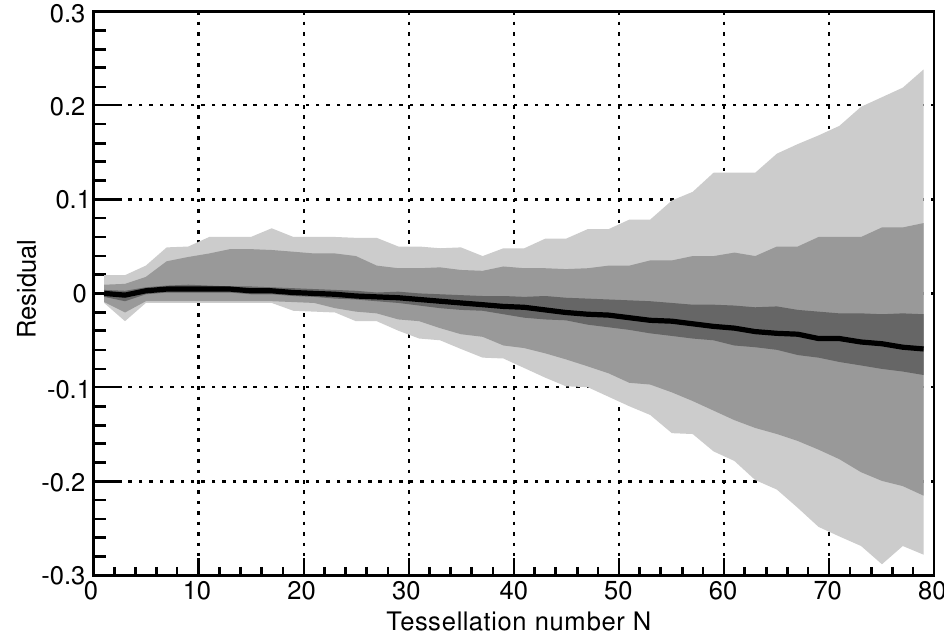}
 \hfill
 \includegraphics*[width=0.48\textwidth,angle=0,clip]{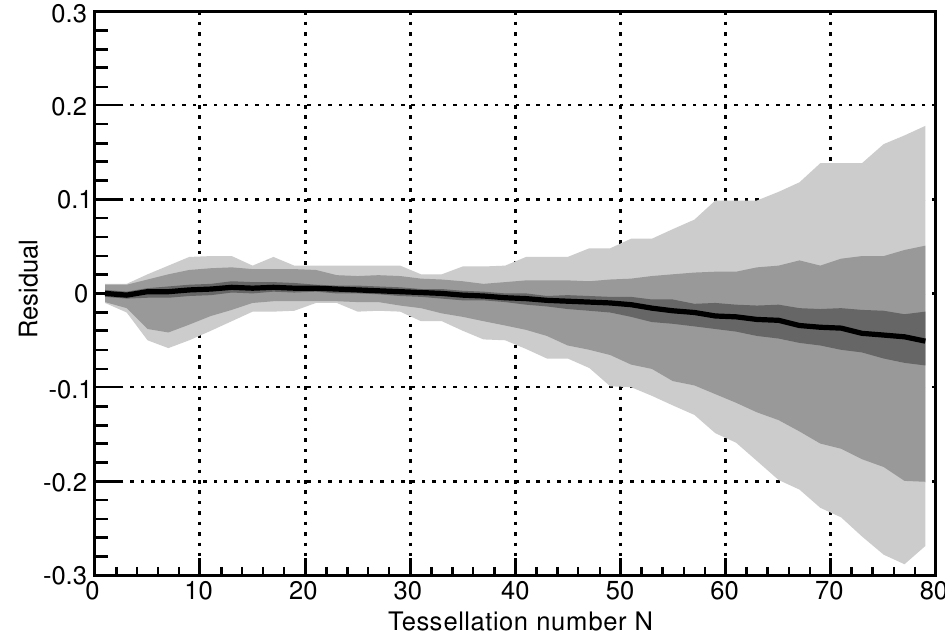}
\caption{Relative residual between the result of Eqs.~\ref{eq:realDC} and the
simulated point-spread function after application of the correction
factor \(w\) in the range \(F/D\)\,=\,[1,2] and \(\alpha\)=[1,6.5].
The black lines corresponds to the median of the distribution. The
gray shaded areas to 68\%, 95\%, and 100\% of the distribution.
The higher deviations at high tessellation ratios are located around
small field-of-views (\(\alpha\)\,\(<\)\,3.5\textdegree{}) and
small \(F/D\lesssim 1.3\). The left and the right plot show the
sagittal and the tangential component, respectively.
\label{fig:Residual}}
\end{figure*}
describing the transition from a single spherical mirror to an ideal
Davies-Cotton reflector. The rescaling factor \(w\) can be interpreted as
the deviation of the shape from the ideal
case. Its value was determined by minimizing the residual, \ie
\(\chi^2\), between simulated point-spread function and approximated
root-mean-square. The differences of the sagittal and tangential
residual are minimized independently for each \(N\). Its value is
depicted in Fig.~\ref{fig:correction}. The introduction of this scale
factor effectively reduced the residual from a maximum of 12\% to less
than 5\% for tessellation numbers smaller than 40. Fig.~\ref{fig:Residual}
shows the distribution of the residuals for different tessellation numbers.

Note that for the case \(N=1\) the simulated single mirror is of hexagonal
shape while the analytical approximation describes a disc-like mirror.
For cases \(N\gg1\) the properties of the simulated reflector 
converge to the ideal Davies-Cotton. While the presented development
was calculated for a disc shaped reflector, here, the simulated
Davies-Cotten converge to an ideal hexagon. Consequently, in both 
cases the rescaling factor is expected to be different from
unity.

In general, it is not expected to obtain a perfect match between 
the analytical approximation and the simulation, because
simulations will always take into account effect which cannot be easily
described analytically, like rays lost between individual mirrors.


From Eqs.~\ref{eq:realDC} it is evident that for rays with small
incident angles the point-spread function is dominated by the \nth{0}-order term which
decreases fast with high tessellation number. At higher incident angles
the point-spread function is dominated by higher order terms which only
turn from the spherical to the ideal Davies-Cotton solution for
increasing tessellation numbers. In general the dominating term for
reasonable incident angles is the \nth{0}-order term. Consequently,
the point-spread function dramatically improves for \(N\)\,\(>\)\,1
but for \(N\)\,\(\ge\)\,5 changes become unimportant. Hence, for
practical purposes a single mirror and the case of \(N=3\) can be
excluded while for most practical purposes \(N=5\) will already be
enough.

\subsection{Summary}

It is possible to describe the optical quality of a set of well defined
Davies-Cotton reflectors quite well in a single analytical formula. 
Even the real Davies-Cotton might be slightly different, \eg
different mirror or reflector shapes or obscuration by the focal plane
instrumentation, this gives a very good estimate of the optical performance.

\end{document}